\documentclass[a4paper,11pt]{article}
\pdfoutput=1 

\usepackage{jheppub}
\usepackage[T1]{fontenc} 

\usepackage[dvipsnames]{xcolor}

\usepackage{tikz}
\usepackage{pgfplots}
\pgfplotsset{compat=newest}
\usetikzlibrary{plotmarks,calc}
\usetikzlibrary{arrows}

\tikzstyle arrow=[thick,rounded corners=18pt,-latex]
\tikzstyle box=[draw,rounded corners,outer sep=4pt]
\tikzstyle B node=[outer sep=0pt]
\tikzstyle Q node=[inner sep=1pt,outer sep=0pt]

\usepackage{enumitem}

\newcommand{\bei}{\begin{itemize}}
\newcommand{\eei}{\end{itemize}}
\newcommand{\bee}{\begin{enumerate}}
\newcommand{\eee}{\end{enumerate}}
\newcommand{\beeL}{\begin{enumerate}[label=(\Alph*)]}
\newcommand{\beel}{\begin{enumerate}[label=(\alph*)]}
\newcommand{\beeR}{\begin{enumerate}[label=(\Roman*)]}
\newcommand{\beer}{\begin{enumerate}[label=(\roman*)]}
\newcommand{\beeLd}{\begin{enumerate}[label=\Alph*.]}
\newcommand{\beeld}{\begin{enumerate}[label=\alph*.]}
\newcommand{\beeRd}{\begin{enumerate}[label=\Roman*.]}
\newcommand{\beerd}{\begin{enumerate}[label=\roman*.]}

\newcommand{\gen}[1]{\mathbf{#1}}
\newcommand{\genr}[1]{\widetilde{\mathbf{#1}}}

\newcommand{\ad}{\dot{\alpha}}
\newcommand{\bd}{\dot{\beta}}

\newcommand{\Ad}{\dot{A}}
\newcommand{\Bd}{\dot{B}}

\newcommand{\su}{\mathfrak{su}}
\newcommand{\psu}{\mathfrak{psu}}

\newcommand{\de}{\text{d}}

\newcommand{\bal}{\begin{equation}\begin{aligned}}
\newcommand{\eal}{\end{aligned}\end{equation}}

\definecolor{grey}{rgb}{0.4,0.4,0.5}
\definecolor{darkgreen}{rgb}{0,0.5,0}
\definecolor{darkred}{rgb}{0.6,0.0,0}
\definecolor{lightbrown}{rgb}{1,0.9,0.8}
\definecolor{brown}{rgb}{0.6,0.3,0.3}
\definecolor{darkblue}{rgb}{0,0,0.5}
\definecolor{darkmagenta}{rgb}{0.5,0,0.5}

\newcommand{\adsst}{${\rm  AdS}_3\times {\rm S}^3\times {\rm T}^4\ $}

\def\a {\alpha}

\renewcommand{\L}{{\scriptscriptstyle\text{L}}}
\newcommand{\R}{{\scriptscriptstyle\text{R}}}

\newcommand{\vac}[1]{\,|0\rangle_{#1}}

\title{\boldmath Comments on Integrability in the Symmetric Orbifold}

\author[a]{Sergey Frolov,%
}
\author[b,c,1]{Alessandro Sfondrini\note{MATRIX Simons Fellow.}}

\affiliation[a]{School of Mathematics and Hamilton Mathematics Institute,\\
Trinity College, Dublin 2, Ireland}
\affiliation[b]{Dipartimento di Fisica e Astronomia, Universit\`a degli Studi di Padova,\\
via Marzolo 8, 35131 Padova, Italy}
\affiliation[c]{Istituto Nazionale di Fisica Nucleare, Sezione di Padova,\\
via Marzolo 8, 35131 Padova, Italy}

\emailAdd{frolovs@maths.tcd.ie}
\emailAdd{alessandro.sfondrini@unipd.it}

\abstract{We present a map between the excitation of the symmetric-product orbifold CFT of $T^4$, and of the worldsheet-integrability description of $AdS_3\times S^3\times T^4$ of Lloyd, Ohlsson Sax, Sfondrini, and Stefa\'nski at $k=1$. We discuss the map in the absence of RR fluxes, when the theory is free, and at small RR flux, $h\ll 1$, where the symmetric-orbifold CFT is deformed by a marginal operator from the twist-two sector.
We discuss the recent results of Gaberdiel, Gopakumar, and Nairz, who computed from the perturbed symmetric-product orbifold the central extension to the symmetry algebra of the theory and its coproduct. We show that it coincides with the $h\ll 1$ expansion of the lightcone symmetry algebra known from worldsheet integrability, and that hence the S~matrix found by Gaberdiel, Gopakumar, and Nairz maps to the one bootstrapped by the worldsheet integrability approach. 
}

\begin{document} 
\maketitle
\flushbottom

\section{Introduction and summary}

Strings on \adsst can be supported by a mixture of Ramond-Ramond (RR) and Neveu-Schwarz-Neveu-Schwarz fluxes (NSNS), see for instance~\cite{Demulder:2023bux} for a recent review. In the case where the RR flux vanishes, the worldsheet model is a supersymmetric Wess-Zumino-Witten model, with level~$k=1,2,3,\dots$. A very special case is that of $k=1$; in that case, the string model is dual~\cite{Eberhardt:2018ouy} to a particularly simple (and well-studied, see e.g.~\cite{Arutyunov:1997gi,Lunin:2000yv}) CFT,  the symmetric product orbifold CFT of $T^4$. This theory has $\mathcal{N}=(4,4)$ superconformal symmetry,%
\footnote{The $\mathcal{N}=(4,4)$ superconformal algebra is infinite-dimensional, contains two copies of the Virasoro algebra, and has as globally-defined subalgebra the Lie superalgebra~$\psu(1,1|2)\oplus\psu(1,1|2)$ which we will often indicate as $\psu(1,1|2)^{\oplus2}$.
}
and is obtained by taking $N$ copies of a free supersymmetric model with central charge $c=6$ and quotienting by the action of the symmetric group~$S_N$. The limit of free strings corresponds to $N\to\infty$. The states of this theory are labeled by the conjugacy classes of the symmetric group, \textit{i.e.}\ by cycles. At large-$N$, it is sufficient to study single-cycle states, which are dual to perturbative closed-string states --- quite similarly to how single-trace operators in $\mathcal{N}=4$ supersymmetric Yang-Mills theory (SYM) are dual to string states.
More specifically, given a sector with a single cycle of length~$w$, states can be constructed by acting freely with fractionally-moded oscillators, such as (for a boson) $\alpha_{-n/w}$ with $n>0$, or $\tilde{\alpha}_{-n'/w}$, again with $n'>0$. This is due to the presence of a twist-$w$ field~$\sigma_w$ in this sector.
The total dimension of the state is then given by the sum of the mode numbers, $(\sum_i n_i+\sum_j n'_j)/w$, plus of course the dimension of $\sigma_w$ (which depends on~$w$).
Not all states so constructed are however physical. Invariance under the orbifold action dictates that the total mode number should be integer, $\sum_i n_i-\sum_j n'_j=0$ $\text{mod}w$ (again, for bosons). This is the analogue of the level-matching condition in string theory. In this sense, it is tempting to think of the oscillators of $\alpha_{-n/w}$ and $\tilde{\alpha}_{-n'/w}$ as analogues of the ``magnons'' which appear in $\mathcal{N}=4$ SYM~\cite{Berenstein:2002jq,Minahan:2002ve}, with momentum $p=2\pi n/w$ or $p'=-2\pi n'/w$ (as they are left- and right-movers).

At $k=1$, $N=\infty$, and without any RR flux, the theory is essentially free (similarly to free planar $\mathcal{N}=4$ SYM). One way to introduce interactions is to turn on the RR flux, whose strength we indicate with~$h$; in perturbative string theory this can be done adiabatically, see e.g.~\cite{OhlssonSax:2018hgc} for a discussion of the identification of the moduli.
From the point of view of the symmetric-product orbifold, this can be done by deforming the theory by means of a marginal operator constructed out of the twist-two sector,
\begin{equation}
    h\,\int\de^2z\, \mathcal{O}_{(2)}(z,\bar{z})\,,\qquad h\ll1\,.
\end{equation}
This perturbation has been extensively studied in the literature see for instance~\cite{Apolo:2022fya,Hughes:2023fot} and references therein, especially in the context of the computation of the anomalous dimension of single-cycle operators.
While for any~$w$ there are a handful of states which are BPS, and do not receive any correction to their scaling dimension, one expects most states to have an anomalous dimensions. More precisely, in the perturbed theory one expects the dimensions \textit{i.e.}\ the eigenvalue of $\gen{L}_0+\genr{L}_0$, to be corrected starting at order~$O(h^2)$. Conversely, the spin $\gen{L}_0-\genr{L}_0$ is quantised and cannot be corrected.
This fact has an interesting consequence. If we consider a BPS state, and act with a few chiral oscillators $\alpha_{-n/w}$ on it (but without any antichiral ones), its dimension under $\genr{L}_0$ must nonetheless change. Because  $\genr{L}_0$ can be expressed as the anti-commutator of $\mathcal{N}=(4,4)$ supercharges, we therefore have that a state containing only chiral oscillators cannot be annihilated by the antichiral supersymmetry generators $\genr{G}\in\psu(1,1|2)$. Already in 2002~\cite{Gomis:2002qi,Gava:2002xb}, it was suggested that this must result in an action of the antichiral generators on the chiral oscillators (and vice-versa),
\begin{equation}
\label{eq:chiralantichiralact}
    \big[\gen{G},\tilde{\alpha}_{-\frac{n_1}{w}}\dots\tilde{\alpha}_{-\frac{n_r}{w}}\big]\neq0\,,\quad
    \big[\genr{G},\alpha_{-\frac{n_1}{w}}\dots\alpha_{-\frac{n_r{}'}{w}}\big]\neq0\,,\qquad\text{in the perturbed theory.}
\end{equation}
Clearly, this action is not compatible with the original $\psu(1,1|2)$ action. Therefore, it must vanish both if $h=0$ or if $\alpha_{-{n_1}/{w}}\dots\alpha_{-{n_r'}/{w}}$ defines by itself a physical state of the theory.

This action is very reminiscent of the central extension that plays an important role in integrability of $\mathcal{N}=4$ SYM~\cite{Beisert:2005tm}, see~\cite{Beisert:2010jr,Arutyunov:2009ga} for review, and in \adsst~\cite{Borsato:2012ud}, see~\cite{Demulder:2023bux,Sfondrini:2014via} for reviews.
In fact, \adsst is believed to be integrable in the planar limit, for arbitrary values of the RR flux~$h$ and of the NSNS flux~$k$. This has been proven to be so at the level of the classical non-linear sigma model action~\cite{Cagnazzo:2012se}.
At the quantum level, integrability manifest itself on the string worldsheet, when the theory is suitably gauge-fixed (in a particular lightcone gauge~\cite{Arutyunov:2005hd}). The gauge-fixing is done with reference to a BPS state, so that half of the supersymmetries of the model survive in lightcone gauge. The theory becomes especially simple in the decompactification limit, where the size of the worldsheet~$w$%
\footnote{It is not a coincidence that we use the same letter for the length of a $w$-cycle and the size of the worldsheet; indeed, in the case that we discuss in this paper, the reference BPS state on the worldsheet is dual to a BPS state in the $w$-twist sector of the symmetric-product orbifold.}
(which is now related to the R-charge of the reference BPS state) is taken to infinity. Then, for any $h,k$ we can study the worldsheet theory on a plane; this allows one to define asymptotic states, of the type 
\begin{equation}
    A^\dagger(p_1)\dots A^\dagger(p_r)|0\rangle_w\,,
\end{equation}
and an S~matrix which scatters them.
In fact, this S-matrix is almost entirely fixed by the symmetries of the lightcone gauge up to a handful of scalar pre-factors, the so-called dressing factors. For the case of generic $h,k$, this S-matrix was fixed in~\cite{Lloyd:2014bsa}.
An instrumental tool in fixing it was precisely the appearance (in the lightcone gauge, and when the level-matching constraint is not imposed) of symmetries of the type~\eqref{eq:chiralantichiralact}, which results in a particular central extension of the lightcone symmetry algebra, so that $\{\gen{G},\genr{G}\}\neq 0$.

It is worth stressing that the study of the symmetries and of the S~matrix in the decompactification limit $w\to\infty$ goes well beyond the study of the BMN limit of the model~\cite{Berenstein:2002jq}. For instance, in the decompactification limit, $k$ and $h$ may take any value, and the energy of a single particle $A^\dagger(p)$ takes the rather non-trivial form~\cite{Hoare:2013lja,Lloyd:2014bsa}
\begin{equation}
\label{eq:dispersion1}
    H(p)=\sqrt{\left(\mu+\frac{k}{2\pi} p\right)^2+4h^2\sin^2\left(\frac{p}{2}\right)}\,,
\end{equation}
where $\mu\in\mathbb{Z}$ is a mass of sorts. Conversely, in the BMN limit one must take the string tension $T=\sqrt{h^2+k^2/(4\pi^2)}$ to be large, and the worldsheet momentum~$p$ to be small, washing away much of the physics of the model.%
\footnote{In particular, it is impossible to consider $k=1$ and $h\ll1$ in the BMN limit, as $T\to\infty$ requires either $k\to\infty$ or to expand in~$k/h$ around $h=\infty$.}

It is natural to ask whether the predictions of integrability, such as~\eqref{eq:dispersion1}, may be derived not from the worldsheet, but directly from the CFT dual, which of course was a big factor in the success of the integrability program for $\mathcal{N}=4$ SYM.
Historically, one of the first attempts to study integrability in AdS${}_3$/CFT$_{2}$ was motivated indeed by the study of the symmetric-product orbifold theory~\cite{David:2008yk,David:2010yg} (see also~\cite{Pakman:2009mi}), though at the time this was believed to be directly related to the RR backgrounds emerging from the D1-D5 system. In turn, these investigations spurred the study of integrability on the sigma-model side of the duality, which was initiated for RR backgrounds in~\cite{Babichenko:2009dk}.
The question of integrability in the deformed symmetric-product orbifold was recently revisited by~\cite{Gaberdiel:2023lco}, building on the early work of~\cite{Gomis:2002qi,Gava:2002xb,David:2010yg}. There, working in first-order perturbation theory around the symmetric product orbifold, \textit{i.e.}\ at 
\begin{equation}
    k=1,\qquad h\ll1\,,
\end{equation}
the authors establish the form of the central extension appearing in the lightcone symmetry algebra. More specifically, through a computation in first-order conformal perturbation theory, the authors determine the action of
\begin{equation}
    \big[\gen{G},\tilde{\alpha}_{-\frac{n_1}{w}}\tilde{\alpha}_{-\frac{n_2}{w}}\big]=h \tilde{\Psi}(n_1,n_2,w) + O(h^2)\,,\quad
    \qquad\text{in the perturbed theory,}
\end{equation}
where $\tilde{\Psi}(n_1,n_2,w)$ is a linear combination of one fermion and one boson oscillator from the undeformed theory, with mode numbers $n_1$ and $n_2$, or $n_2$ and $n_1$. The coefficients of the linear combination are explicit functions of   $(n_1,n_2,w)$, and they reproduce all the characteristics that one expects from integrability, including vanishing for physical values of $n_1,n_2$, having a well defined planar limit $w\to\infty$.
The authors then use the form of these coefficients in the decompactification limit to infer the form of the central extension; because the central extension is linear in~$h$, they reproduce the algebra of~\cite{Lloyd:2014bsa}, including its non-trivial coproduct (which they can also extract from the two-particle representation). They also compute an S~matrix using that symmetry algebra, and compare their finding with the BMN limit of the theory.

In this paper we make more precise the relation of the results of~\cite{Gaberdiel:2023lco} with the existing integrability literature. In particular, after reviewing the salient points of the integrability construction, we argue what follows.
\begin{enumerate}
    \item The correspondence between the modes of symmetric-product orbifold and of the string should be established in the decompactification limit (at $k=1$ and $h\ll1$) rather than in the BMN limit. The map is provided in eqs.~\eqref{eq:chiralident} and~\eqref{eq:antichiralident}. Indeed in the BMN limit the tension $T=\sqrt{\tfrac{k^2}{4\pi^2}+h^2}$ goes to infinity with $k/(2\pi h)$ fixed, which makes it impossible to distinguish  the $k=1$ case from the pure-RR one.
    \item As was argued in~\cite{Frolov:2023lwd}, if we allow the momentum $p$ to take arbitrary real values (rather than restricting to $0\leq p< 2\pi$), at $k=1$ we should only consider states with $\mu=0$ (and more generally, $|\mu|\leq k-1$). However, again as argued in~\cite{Frolov:2023lwd}, one can consider higher-$|\mu|$ states as bound states if one restricts the value of~$p$.
    \item The authors find the representation of the supercharges at linear order in their deformation parameter~$g$, and use it to fix the central charge. Through a shortening condition, this gives~\eqref{eq:chiralantichiralact}. However, this does not exclude that the function $h(g)$ has a non-trivial expansion of the type~$h=g+c_1g^3+\dots$, but it is just a consequence of the form of the shortening condition and of the fact that the central extension appears at linear order in~$h$.
    \item Because the authors determine the S~matrix by a symmetry bootstrap like the one of~\cite{Lloyd:2014bsa}, it is sufficient to compare the representations derived  at linear order of the perturbation in~\cite{Gaberdiel:2023lco} with the expansion of the worldsheet representations of~\cite{Lloyd:2014bsa} in the case $k=1$, $h\ll 1$. By doing so, we find that~\cite{Gaberdiel:2023lco} indeed reproduces a limit of the construction of~\cite{Lloyd:2014bsa}, and therefore that the two S~matrices must be equivalent. We also derive the relation between them, which is a little involved just because the two representations have different co-products.%
    \footnote{More specifically,  the construction of~\cite{Lloyd:2014bsa} is in the ``string frame'' and that of~\cite{Gaberdiel:2023lco} in the ``spin-chain frame''; such small discrepancies are also well-known in $\mathcal{N}=4$ SYM~\cite{Arutyunov:2006yd}.}
\end{enumerate}
All in all, this shows that the results of~\cite{Gaberdiel:2023lco} fit perfectly within the existing integrability construction~\cite{Lloyd:2014bsa}, and therefore that direct computation in conformal perturbation theory in the symmetric-product orbifold CFT should provide new insight into completing the integrability program for this duality. We very much hope that this will allow the community to reach the resounding successes obtained in the study of $\mathcal{N}=4$ SYM and $AdS_5\times S^5$ integrability.

\section{A brief review of worldsheet integrability}
\subsection{Symmetries}
The Killing spinors of the \adsst close in two copies of the $\mathfrak{psu}(1,1|2)$ superalgebra. 
When it will be necessary to distinguish the two copies, we will indicate with ``left'' and ``right'', and write
\begin{equation}
    \psu(1,1|2)_{\L}\oplus\psu(1,1|2)_{\R}\,.
\end{equation}
From the torus directions we also have four $\mathfrak{u}(1)$ shift isometries as well as, locally, an $\mathfrak{so}(4)$ symmetry, which we decompose as 
\begin{equation}
   \mathfrak{so}(4)_{\text{T}^4}\cong\mathfrak{su}(2)_\bullet\oplus\mathfrak{su}(2)_\circ\,, 
\end{equation}
with generators
$\gen{J}_{\bullet}{}^{AB}$ and $\gen{J}_{\circ}{}^{\dot{A}\dot{B}}$,
where $A,B$ and $\dot{A},\dot{B}$ may take values $1,2$ and $\dot{1}, \dot{2}$, respectively.
The anti-symmetric combination of the indices $A,B$  in $\gen{J}_{\bullet}{}^{AB}$ decouples and is not part of $\mathfrak{su}(2)_{\bullet}$, and similarly for $\dot{A},\dot{B}$.
Both set of generators satisfy the standard $\mathfrak{su}(2)$ relations.

Let us now consider one of the two copies of $\mathfrak{psu}(1,1|2)$, let's say $\mathfrak{psu}(1,1|2)_{\L}$. Its bosonic subalgebra is
\begin{equation}
    \mathfrak{sl}(2,\mathbb{R})_{\L}\oplus\mathfrak{su}(2)_{\L}
    \cong \mathfrak{su}(1,1)_{\L}\oplus\mathfrak{su}(2)_{\L}
    \subset \mathfrak{psu}(1,1|2)_{\L}\,.
\end{equation}
This has generators $\gen{L}_{n}$ with $n=0,\pm1$ for $\mathfrak{sl}(2,\mathbb{R})_{\L}$, and $\gen{J}^{\alpha\beta}$ with $\alpha,\beta=\pm$ for $\mathfrak{su}(2)_{\L}$. Once again anti-symmetric combination of $\alpha\beta$ in $\gen{J}^{\alpha\beta}$ decouples. From $\mathfrak{psu}(1,1|2)_{\R}$ we have the bosonic subalgebras $\mathfrak{sl}(2,\mathbb{R})_{\R}$  and $\mathfrak{su}(2)_{\R}$. We will denote the generators of the latter as $\genr{J}^{\ad\bd}$, where the indices may take values $\dot{\pm}$.
We summarise the notation for the four $\su(2)$ algebras in Table~\ref{tab:su2}.

\begin{table}[t]
\centering
 \begin{tabular}{|c|c c |p{8cm}|} 
 \hline
 Algebra & Generators & Indices & Interpretation \\ [0.5ex] 
 \hline
 $\su(2)_\bullet$ & $\gen{J}_{\bullet}{}^{AB}$ & $A,B=1,2$ & From the torus $\mathfrak{so}(4)\cong\mathfrak{su}(2)_\bullet\oplus\mathfrak{su}(2)_\circ$. We will see below that it acts as an automorphism on $\psu(1,1|2)_{\L}$ and on $\psu(1,1|2)_{\R}$.\\ 
$\su(2)_\circ$ & $\gen{J}_{\circ}{}^{\Ad\Bd}$ & $\Ad,\Bd=\dot{1},\dot{2}$ & From the torus $\mathfrak{so}(4)\cong\mathfrak{su}(2)_\bullet\oplus\mathfrak{su}(2)_\circ$. We will see below that it commutes with $\psu(1,1|2)_{\L}$ and with $\psu(1,1|2)_{\R}$. \\
$\su(2)_{\L}$ & $\gen{J}^{\alpha\beta}$ & $\alpha,\beta=\pm$ & ``Left'' R-symmetry, $\su(2)_{\L}\subset\psu(1,1|2)_{\L}$. \\
$\su(2)_{\R}$& $\genr{J}^{\ad\bd}$ & $\ad,\bd=\dot{\pm}$ & ``Right'' R-symmetry, $\su(2)_{\R}\subset\psu(1,1|2)_{\R}$. \\
 \hline
 \end{tabular}
 \label{tab:su2}
 \caption{Summary of the notation for the four $\su(2)$ algebras appearing in \adsst.}
\end{table} 

Let us describe in more detail the (anti)commutation relations which we will need later.
For $\gen{sl}(2,\mathbb{R})$ we have
\begin{equation}
    [\gen{L}_{n},\gen{L}_{m}]=(n-m)\gen{L}_{n-m}\,.
\end{equation}
In the dual CFT, $\gen{L}_{-1}$ is the generator of (chiral) translations, while $\gen{L}_0$ corresponds to dilatations and $\gen{L}_{+1}$ to special conformal transformations.
For $\mathfrak{su}(2)$ we write, in the basis $\gen{J}^{\pm\pm}$ and $\gen{J}^3=\frac{1}{2}(\gen{J}^{+-}+\gen{J}^{-+})$,
\begin{equation}
    [\gen{J}^3,\gen{J}^{\pm\pm}]= \pm\gen{J}^{\pm\pm}\,,\qquad
    [\gen{J}^{++},\gen{J}^{--}]=2\gen{J}^3\,.
\end{equation}
In the dual CFT these represent the chiral part of the R-symmetry algebra.

The superalgebra $\mathfrak{psu}(1,1|2)$ has eight odd generators. Four of them have the interpretation of supersymmetry generators in the dual CFT, and we indicate them with $\gen{Q}^{\alpha A}$, where $A=1,2$ is the index of the fundamental (or ``\textbf{2}'') representation of $\mathfrak{su}(2)_{\bullet}$; they have positive scaling dimension. The remaining four, which we call $\gen{S}^{\alpha A}$,
have the interpretation of superconformal generators, with negative scaling dimension. We have
\begin{equation}
\begin{aligned}
 \relax    [\gen{L}_0,\gen{Q}^{\alpha A}] &=
    +\frac{1}{2}\gen{Q}^{\alpha A}\,,\qquad&&
    [\gen{L}_0,\gen{S}^{\alpha A}] =
    -\frac{1}{2}\gen{S}^{\alpha A}\,,\\
    [\gen{L}_{+1},\gen{Q}^{\alpha A}]&=+\gen{S}^{\alpha A},\qquad &&
    [\gen{L}_{-1},\gen{S}^{\alpha A}]=-\gen{Q}^{\alpha A}.
\end{aligned}
\end{equation}
All odd generators carry a fundamental R-symmetry $\mathfrak{su}(2)$ index, and obey the standard relations
\begin{equation}
    [\gen{J}^3,\gen{Q}^{\pm A}]=\pm\frac{1}{2}\gen{Q}^{\pm A}\,,\qquad
    [\gen{J}^{\pm\pm},\gen{Q}^{\mp A}]=\gen{Q}^{\pm A}\,,
\end{equation}
and similarly for~$\gen{S}^{\alpha A}$.
Note that these generators are also charged under $\mathfrak{su}(2)_{\bullet}$ (due to the index~$A$). In fact, $\mathfrak{su}(2)_{\bullet}$ is an outer automorphism of the $\mathfrak{psu}(1,1|2)$ algebra. This can be seen from the anti-commutation relations. The non-vanishing ones are
\begin{equation}
\begin{aligned}
    \{\gen{Q}^{\alpha A},\gen{Q}^{\beta B}\}&=
    \varepsilon^{\alpha\beta}\varepsilon^{AB}\,2\gen{L}_{-1}\,,
    \qquad&&
    \{\gen{S}^{\alpha A},\gen{S}^{\beta B}\}=
    \varepsilon^{\alpha\beta}\varepsilon^{AB}\,2\gen{L}_{+1}\,,\\
    \{\gen{Q}^{\pm, A},\gen{S}^{\pm, B}\}&=
    \pm\varepsilon^{AB}\,2\gen{J}^{\pm\pm}\,,\qquad&&
    \{\gen{Q}^{\pm, A},\gen{S}^{\mp, B}\}=
    \pm\epsilon^{AB}\,2\left(\gen{L}_0-\gen{J}^3\right)\,.
\end{aligned}
\end{equation}

The second copy of $\mathfrak{psu}(1,1|2)$ is related to the anti-chiral excitations of the dual CFT. It is given by the generators
\begin{equation}
    \genr{L}_{\dot{n}},\qquad \genr{J}^{\dot{\alpha}\dot{\beta}},\qquad\genr{Q}^{\dot{\alpha}A},\qquad \genr{S}^{\dot{\alpha}A},\qquad
    \text{with}\quad
    \dot{n}=0,\pm1,\quad \dot{\alpha},\dot{\beta}=\pm,\quad
    A= 1,2\,.
\end{equation}
or more precisely $\dot{n}=\dot{0},\pm\dot{1},\quad \dot{\alpha},\dot{\beta}=\dot{\pm}$ (we will often suppress the dots on the indices to avoid cluttering the notation).

They satisfy identical commutation relations. Notice that this anti-chiral copy of $\mathfrak{psu}(1,1|2)$ is charged under the same $\mathfrak{su}(2)_{\bullet}$. Notice also that both copies of $\mathfrak{psu}(1,1|2)$ commute with $\mathfrak{su}(2)_{\circ}$. We stress that all chiral generators (anti)commute with the anti-chiral ones,
\begin{equation}
\begin{aligned}
    &[\gen{L}_{n},\genr{L}_{\dot{n}}]=
    [\gen{L}_{n},\genr{Q}^{\dot{\alpha} A}]=
    [\gen{L}_{n},\genr{S}^{\dot{\alpha} A}]=
    [\gen{J}^{\alpha\beta},\genr{J}^{\dot{\alpha}\dot{\beta}}]=\dots &=&\ 0\,,\\
    &\{\gen{Q}^{\alpha A},\genr{Q}^{\dot{\alpha} B}\}=
    \{\gen{Q}^{\alpha A},\genr{S}^{\dot{\alpha} B}\}=
    \{\gen{S}^{\alpha A},\genr{Q}^{\dot{\alpha} B}\}=
    \{\gen{S}^{\alpha A},\genr{S}^{\dot{\alpha} B}\}&=&\ 0\,.
\end{aligned}
\end{equation}
Note finally that in the CFT literature the supersymmetry and superconformal generators are often indicated as
\begin{equation}
\label{eq:superchargesnotation}
    \gen{G}_{-\frac{1}{2}}^{\alpha A}=\gen{Q}^{\alpha A}\,,\qquad
    \gen{G}_{+\frac{1}{2}}^{\alpha A}=\gen{S}^{\alpha A}\,,\qquad
    \genr{G}_{-\frac{1}{2}}^{\dot\alpha A}=\genr{Q}^{\dot\alpha A}\,,\qquad
    \genr{G}_{+\frac{1}{2}}^{\dot\alpha A}=\genr{S}^{\dot\alpha A}\,.
\end{equation}

\subsection{Light-cone gauge fixing and residual symmetries}
\label{sec:review:lcgauge}
The left and right $\psu(1,1|2)$ algebras have the BPS bounds
\begin{equation}
    \gen{L}_0-\gen{J}^3\geq0\,,\qquad
    \genr{L}_0-\genr{J}^3\geq0\,.
\end{equation}
Highest-weight states which saturate one of these bounds correspond to left or right short representations. In that case, the highest-weight state is annihilated  by half of the lowering operators. In the context of the integrability construction, it is particularly convenient to consider a reference highest-weight state~$\vac{w}$ which is short with respect to \textit{both} the left and right algebras, so that
\begin{equation}
\label{eq:halfBPSstate}
    \gen{L}_0\vac{w}=\gen{J}^3\vac{w}=\genr{L}_0\vac{w}=\genr{J}^3\vac{w}=\frac{w}{2}\vac{w}\,,\qquad
    w\in\mathbb{Z}\,.
\end{equation}
This is \textit{half-BPS}, meaning that it is annihilated by half of the supercharges of $\psu(1,1|2)_{\L}\oplus \psu(1,1|2)_{\R}$, \textit{i.e.}\ by eight real supercharges. The algebra preserved by this choice will play an important role in the construction of the worldsheet S~matrix.

Let us review how the choice~\eqref{eq:halfBPSstate} comes up in the light-cone gauge construction of the string sigma model~\cite{Borsato:2014hja}. We introduce lightcone coordinates and conjugate momenta
\begin{equation}
\label{eq:lcgauge}
    x^\pm =\frac{\phi\pm t}{2}\,,\qquad
    p_{\pm} = p_\phi\pm p_t\,,
\end{equation}
where $t$ is time in $\text{AdS}_3$ and $\phi$ is a great circle on~$\text{S}^3$. The gauge-fixing condition is then~\cite{Lloyd:2014bsa}
\begin{equation}
    x^+ = \tau\,,\qquad p_- = 2\,,
\end{equation}
where $\tau$ is the worldsheet time and the constant~$2$ is chosen for a convenient normalisation.
As~$x^+$ becomes related to worldsheet time, the conjugate momentum~$p_+$ becomes related to the Hamiltonian density of the gauge-fixed model. This allows us to identify the worldsheet Hamiltonian as
\begin{equation}
    \gen{H}=(\gen{L}_0-\gen{J}^3)+(\genr{L}_0-\genr{J}^3)\geq 0\,.
\end{equation}
The combination of the Cartan generators related to the integral of $p_-$ also takes a special meaning: since the density $p_-$ is constant, it becomes related to the classical circumference~$r$ of the worldsheet. In particular,  we have
\begin{equation}
    2r = \int\limits_{0}^{r}\de\sigma\, p_- = \gen{L}_0+\gen{J}^3+\genr{L}_0+\genr{J}^3\,,
\end{equation}
so that we define
\begin{equation}
    \gen{R}=\frac{1}{2}\left(\gen{L}_0+\gen{J}^3+\genr{L}_0+\genr{J}^3\right)\,.
\end{equation}
Note that on our reference half-BPS state
\begin{equation}
    \gen{R}\vac{w}=w\vac{w}\,.
\end{equation}
There are two more combinations of the Cartan charges which we may consider, and we introduce the notation
\begin{equation}
    \gen{M}=(\gen{L}_0-\gen{J}^3)-(\genr{L}_0-\genr{J}^3)\,,
    \qquad
    \gen{B}=(\gen{L}_0+\gen{J}^3)-(\genr{L}_0+\genr{J}^3)\,.
\end{equation}
Both of these charges, as well as the Hamiltonian, annihilate our reference state
\begin{equation}
\gen{H}\vac{w}=
\gen{M}\vac{w}=
\gen{B}\vac{w}=0\,,
\end{equation}
which will become the vacuum of the gauge-fixed model.
It is worth noting that $\gen{M}$ and $\gen{B}$ must take quantised values on physical states, unlike $\gen{H}$. In fact, while $\gen{L}_0+\genr{L}_0$ is the (real) energy, $\gen{L}_0-\genr{L}_0$ is the (quantised) spin.

\subsubsection{Worldsheet momentum and level matching}
Another generator which plays an important role is the worldsheet momentum~$\gen{p}$, which in the classical theory is the Noether charge of $\sigma$-translations on the worldsheet. In the quantum theory, we want to impose the level-matching condition for physical states. For the simplest gauge choice~\eqref{eq:lcgauge} we have
\begin{equation}
    \gen{p}\,|\text{phys}\rangle = 0\,.
\end{equation}
This can be relaxed by allowing winding along the compact light-cone coordinate $\phi$, $\phi(r)-\phi(0)=2\pi n$, with $n\in\mathbb{Z}$ and suitably modifying~\eqref{eq:lcgauge}, see for instance the discussion in~\cite{Arutyunov:2009ga}. In this way we obtain the familiar condition
\begin{equation}
\label{eq:levelmatching}
    \gen{p}\,|\text{phys}\rangle = 2\pi n\,|\text{phys}\rangle\,,\qquad
    n\in\mathbb{Z}\,.
\end{equation}

\paragraph{Symmetries of the light-cone gauge vacuum.}
The sub-algebra of $\psu(1,1|2)_{\L}\oplus \psu(1,1|2)_{\R}$ which annihilates the vacuum~$\vac{w}$ is given by
\begin{equation}
\label{eq:chargesthatannihilate}
    \gen{Q}^{A}\equiv \gen{Q}^{+A}\,,\quad
    \gen{S}_{A}\equiv \varepsilon_{AB}\gen{S}^{-B}\,,\qquad
    \genr{Q}_{A}\equiv \varepsilon_{AB}\genr{Q}^{\dot{+}B}\,,\quad
    \genr{S}^{A}\equiv \gen{S}^{\dot{-}A}\,,\qquad
    \gen{H}\,,\quad\gen{M}\,,\quad\gen{B}\,.
\end{equation}
The supercharges form the algebra $[\su(1|1)^{\oplus2}/\mathfrak{u}(1)]^{\oplus2}$
\begin{equation}
\begin{aligned}
    &\{\gen{Q}^{A},\gen{S}_B\}=\delta^A_B\big(\gen{L}_0-\gen{J}^3\big)=\frac{1}{2}\delta^A_B\big(\gen{H}+\gen{M}\big)\,,\\
    &\{\genr{Q}_{A},\genr{S}^B\}=\delta_A^B\big(\genr{L}_0-\genr{J}^3\big)=\frac{1}{2}\delta_A^B\big(\gen{H}-\gen{M}\big)\,,
\end{aligned}
\end{equation}
on which $\gen{B}$ acts as an automorphism,
\begin{equation}
\label{eq:Bautomorphism}
\begin{aligned}
    &[\gen{B},\gen{Q}^{A}] = +\gen{Q}^{A},\qquad
    &&[\gen{B},\gen{S}_{A}] = -\gen{S}_{A},\\
    &[\gen{B},\genr{Q}_{A}] = -\genr{Q}_{A},\qquad
    &&[\gen{B},\genr{S}^{A}] = +\genr{S}^{A}.
\end{aligned}
\end{equation}
All the generators~\eqref{eq:chargesthatannihilate} also commute with the lightcone Hamiltonian $\gen{H}$, and are thus symmetries of the lightcone gauge-fixed model. The charge $\gen{R}$ plays a special role because, while it commutes with the Hamiltonian, it does not annihilate the vacuum.
This will be important in what follows, when we will consider excitations over the vacuum of the form $|\varPsi\rangle={O}_\varPsi^\dagger\vac{w}$, where ${O}_\varPsi^\dagger$ is some combination of suitably-defined creation operators. For any charge that annihilates the vacuum, such as $\gen{Q}^A$, we have
\begin{equation}
\gen{Q}^A\,|\varPsi\rangle
=
\big[\gen{Q}^A,\,{O}_\varPsi^\dagger\big]\vac{w}\,,
\end{equation}
so that we will fit ${O}_\varPsi^\dagger$ in a representation of the algebra generated by~\eqref{eq:chargesthatannihilate}. The same is not possible with~$\gen{R}$, so that we will not be able to assign a well-defined charge under $\gen{R}$ to the excitations in~${O}_\varPsi^\dagger$. 

\subsubsection{Central extension}
Similarly to what happens in $\text{AdS}_5\times\text{S}^5$~\cite{Beisert:2005tm,Arutyunov:2006yd}, the symmetry algebra acting on the gauge-fixed model is not a  subalgebra of $\psu(1,1|2)_{\L}\oplus \psu(1,1|2)_{\R}$ like~\eqref{eq:chargesthatannihilate}: it is \textit{larger}. In fact, it is given by a twofold central extension of the above algebra, namely
\begin{equation}
\label{eq:centrallyextalg}
\begin{aligned}
    &\{\gen{Q}^{A},\gen{S}_B\}=\frac{1}{2}\delta^A_B\big(\gen{H}+\gen{M}\big)\,,\qquad
    &&\{\genr{Q}_{A},\genr{S}^B\}=\frac{1}{2}\delta_A^B\big(\gen{H}-\gen{M}\big)\,,\\
    &\{\gen{Q}^{A},\genr{Q}_B\} = \delta^A_B\,\gen{C}\,,\qquad
    &&\{\gen{S}_{A},\genr{S}^B\} = \delta_A^B\,\gen{C}^\dagger\,,
\end{aligned}
\end{equation}
where the last line is new. The action of~$\gen{B}$ is unchanged from~\eqref{eq:Bautomorphism}, and clearly the central elements commute with~$\gen{B}$.
The fact that now the left- and right-supercharges have non-trivial anticommutation relations can appear strange from the point of view of the dual conformal field theory, but it is important to stress that this central extension is non-trivial only on \textit{non-physical states}, meaning on \textit{states that do not satisfy the level matching condition}~\eqref{eq:levelmatching}. As long as a state is physical,
\begin{equation}
\label{eq:physicalcond1}
    \gen{C}\,|\text{phys}\rangle= \gen{C}^\dagger|\text{phys}\rangle=0\,,
\end{equation}
as well as 
\begin{equation}
\label{eq:physicalcond2}
    \gen{H}\,|\text{phys}\rangle=H\,|\text{phys}\rangle,\quad
    \gen{M}\,|\text{phys}\rangle=M\,|\text{phys}\rangle\,,\qquad\text{with}\quad H\geq0\,,\quad M\in\mathbb{Z}\,.
\end{equation}
This is very similar to what happens in $\text{AdS}_5\times\text{S}^5$~\cite{Beisert:2005tm,Arutyunov:2006yd}. It is interesting to note that the first sign of the central extension~\eqref{eq:centrallyextalg} was found precisely by Gava and Narain in the pp-wave limit of the \adsst string model~\cite{Gava:2002xb} --- in fact, before this central extension was noted in $\text{AdS}_5/\text{CFT}_4$.

\begin{table}[t]
\centering
\renewcommand{\arraystretch}{1.25}
 \begin{tabular}{|l||c | c| c| c |} 
 \hline
            &$\gen{L}_0$ & $\gen{J}^3$ & $\genr{L}_0$  & $\genr{J}^3$  \\
 \hline
 $\gen{Q}^A$ &   $+\tfrac{1}{2}$   &   $+\tfrac{1}{2}$    &   $0$    &$0$      \\
 $\gen{S}_A$ &   $-\tfrac{1}{2}$  &   $-\tfrac{1}{2}$    &    $0$   &       $0$\\
 \hline
 $\genr{Q}_A$ &   $0$   &   $0$    &   $+\tfrac{1}{2}$    &   $+\tfrac{1}{2}$ \\
 $\genr{S}^A$ &    $0$  &    $0$   &   $-\tfrac{1}{2}$    &  $-\tfrac{1}{2}$    \\
 \hline
 $\gen{C}$&   $+\tfrac{1}{2}$   &  $+\tfrac{1}{2}$    &   $+\tfrac{1}{2}$    &   $+\tfrac{1}{2}$   \\
 $\gen{C}^\dagger$&   $-\tfrac{1}{2}$    &   $-\tfrac{1}{2}$    &   $-\tfrac{1}{2}$    &   $-\tfrac{1}{2}$  \\
 \hline
 \end{tabular}
 \caption{The $\mathfrak{u}(1)$ charges of the generators of the light-cone symmetry algebra, including the central charges  $\gen{C}$ and  $\gen{C}^\dagger$.}
 \label{tab:charges}
\end{table}

\subsubsection{Physical values for the  central charges.}
The last ingredient in understanding the symmetry algebra~\eqref{eq:centrallyextalg} is to relate the value of its central charges to physical quantities such as the coupling constants of the model under consideration and the worldsheet momentum~$\gen{p}$. This was done in~\cite{Lloyd:2014bsa} following~\cite{Hoare:2013lja} (see also~\cite{Borsato:2014hja}). We find
\begin{equation}
\label{eq:centralchargesp}
    \gen{C}=\frac{ih}{2}\left(e^{i\gen{p}}-1\right)e^{2i\xi(\gen{p})}\,,
    \quad
    \gen{C}^\dagger=\frac{h}{2i}\left(e^{-i\gen{p}}-1\right)e^{-2i\xi(\gen{p})}\,,
    \quad
    \gen{M}= \mu+\frac{k}{2\pi}\gen{p}\,,
\end{equation}
where $h\geq 0$ semiclassically is the coefficient of the RR three-form flux in the classical string action, $k\in\mathbb{N}$ is the coefficient of the WZ term, and $\mu$ is a free parameter which will distinguish different representations of the algebra. Notice that the expressions of $\gen{C}$ and $\gen{C}^\dagger$ guarantee that~\eqref{eq:physicalcond1} is satisfied once the level-matching constraint~\eqref{eq:levelmatching} is imposed.
As for~$\gen{M}$, imposing the physical condition \eqref{eq:physicalcond2} and the level-matching ~\eqref{eq:levelmatching} yields that it must be
\begin{equation}
    \mu\in\mathbb{Z}\,.
\end{equation}
The undetermined phase~$\xi(\gen{p})$ is a leftover ambiguity; semiclassically, it originates from the lightcone gauge-fixed model~\cite{Arutyunov:2006ak}. More generally, it can be undone by rescaling the supercharges through a $\mathfrak{u}(1)$ automorphism of the original algebra $\mathfrak{psu}(1,1|2)^{\oplus2}$. In the integrability literature we often set
\begin{equation}
    \xi(\gen{p})=0\,,
\end{equation}
though it is sometimes convenient to set
\begin{equation}
    \xi(\gen{p})=-\frac{1}{4}\gen{p}\,,
\end{equation}
which makes $\gen{C}$ and $\gen{C}^\dagger$ real.

It is worth remarking that while the form of the charges~\eqref{eq:centralchargesp} was determined based on semiclassical arguments, it appears unlikely that it may be modified when going towards weak tension of the string. The only freedom is in the relation of the parameter~$h$ with the string tension, which may be given by a complicated interpolating function like it is the case for ABJM~\cite{Gromov:2014eha}. The parameters $k$ and $\mu$ are quantised and cannot change continuously.

Finally it is important to notice that, while the new central elements $\gen{C}$ and~$\gen{C}^\dagger$ are uncharged under $\gen{H}, \gen{M}$ and $\gen{B}$, they are actually charged under~$\gen{R}$,
\begin{equation}
    [\gen{R},\gen{C}] = +\gen{C}\,,\qquad
    [\gen{R},\gen{C}^\dagger] = -\gen{C}^\dagger\,.
\end{equation}
We collect the charges of the various elements of the light-cone symmetry algebra in Table~\ref{tab:charges}. The interpretation is that $\gen{C}$ and $\gen{C}^\dagger$ are \textit{length-changing} operators, which relate states built out of vacua of different charges, such as $\vac{w}$ and $\vac{w\pm1}$. This action is not immediately visible on the vacuum itself, or on physical states,  which are annihilated by these charges as in~\eqref{eq:physicalcond1}. We will see this action in more detail in the next subsection.

\subsection{Representations in the decompactification limit for \texorpdfstring{$k=1$}{k=1}}
\label{sec:review:reprs}

\subsubsection{The decompactification limit}
The structure of the representations simplifies substantially in the limit where the size of the worldsheet goes to infinity,
\begin{equation}
    w\to\infty\,.
\end{equation}
This can be seen both mathematically and physically. Mathematically, the generator~$\gen{R}$ decouples, and we do not need account for the length-changing effects (the length is strictly speaking infinite). Physically, the worldsheet theory is now defined on a line, rather than on a circle, so that we can consider ``well-separated excitations'' and construct asymptotic \textit{in} and \textit{out} states.%
\footnote{%
This argument becomes a little subtle in the case when these excitations do not have a mass gap, but we will see that the construction is possible in this case too.} 
For $k$ sufficiently large, these excitations are the modes of the $8+8$ transverse modes of the string (as well as bound states thereof). While for generic values of $k$ (in particular, $k\neq1$ and $k\neq2$) the existence of these excitations can be understood order by order in a perturbative expansion around the BMN geodesic, \textit{we stress that we are considering a fully interacting theory in large volume, without taking the near-BMN limit}.
Let us consider the modes corresponding to some excitation on the half-BPS vacuum; for instance, we could take bosonic excitations coming from~$T^4$.
Introducing Zamolodchikov-Faddeev creation and annihilation operators (see e.g.~\cite{Arutyunov:2009ga}) for each of these modes $A^\dagger_{j}(p)$, where $j=A\dot{A}$ denotes the specific mode of~$T^4$ we can create excitations over the vacuum of definite momenta $p_1>p_2>\ldots> p_k$,
\begin{equation}
    A^\dagger_{j_1}(p_1)\cdots A^\dagger_{j_k}(p_k)\vac{\infty} = |X_{j_1}(p_1)\ldots X_{j_k}(p_k)\rangle_{\infty}\,,
\end{equation}
where we ordered the momenta so that the velocities of the particles are $v_1>\dots> v_k$, which gives an \textit{in} state, and the subscript indices distinguish the types of particle. This is an explicit way to realise the state $|\varPsi\rangle$ discussed above, where we have written $O^\dagger_\varPsi = A^\dagger_{j_1}(p_1)\cdots A^\dagger_{j_k}(p_k)$.
This state is physical if it satisfies the level-matching condition,
\begin{equation}
    \sum_{j=1}^k p_j=2\pi n\,,\qquad n\in\mathbb{Z}\,.
\end{equation}
Note that even if the $k$-particle state is physical, any given pair of momenta does not have to satisfy the level-matching condition.
It is believed that this model is integrable so that the $k$-body S~matrix factorises into a sequence of $k(k+1)/2$ two-to-two elastic scattering processes.  
Therefore, the two-to-two S-matrix will be constrained by the centrally extended algebra~\eqref{eq:centrallyextalg}, and it is crucial to understand the one- and two-particle representations of that algebra.

\subsection{Shortening condition on the representations}
The one-particle irreducible representations should be uniquely determined by the eigenvalues of the central charges $H,M, C, C^\dagger$ and  depend on the particle momentum~$p$.
Furthermore, we require that as long as the momentum is real it should be $(C^\dagger)^*=C$ and that $H\geq0$ (in fact, with our convention we straightforwardly have $C=C^\dagger$). Both conditions follow from the reality of the algebra~\eqref{eq:centrallyextalg}. Finally, in the very special case when $p=0\,\text{mod}\,2\pi$ we should recover~\eqref{eq:physicalcond1} and~\eqref{eq:physicalcond2}.
Aside from these very general considerations, the dimensions of the representations which actually occurs in this particular models can be determined from a perturbative analysis. It turns out that all $8+8$ transverse excitations of the string transform in $(2+2)$-dimensional representations (meaning: representations containing two bosons and two fermions), for a total of four such irreducible representations.
It is easy to see that the typical representations of the algebra~\eqref{eq:centrallyextalg} are 16-dimensional. For a representation to be 4-dimensional, it must obey a shortening condition which can be expressed in terms of the central charge as follows~\cite{Borsato:2012ud}:
\begin{equation}
\label{eq:shortening}
    H^2 = M^2 +4\, C\,C^\dagger\,.
\end{equation}

\begin{figure}[t]
\centering
  \begin{tikzpicture}[%
    box/.style={outer sep=1pt},
    Q node/.style={inner sep=1pt,outer sep=0pt},
    arrow/.style={-latex}
    ]%

    \node [box] (PhiM) at ( 0  , 2.5cm) { $Y(p)$};
    \node [box] (PsiP) at (-2.5cm, 0cm) { $\eta^1(p)$};
    \node [box] (PsiM) at (+2.5cm, 0cm) { $\eta^2(p)$};
    \node [box] (PhiP) at ( 0  ,-2.5cm) { $Z(p)$};

    \newcommand{\horshift}{0.09cm,0cm}
    \newcommand{\vershift}{0cm,0.10cm}
 
  \draw [arrow,color=OliveGreen,thick] ($(PhiM.west) +(\vershift)$) -- ($(PsiP.north)-(\horshift)$) node [pos=0.25,anchor=south east,Q node,rotate=45] { $\gen{Q}^1,\genr{S}^1$};
    \draw [arrow,color=Orange,thick] ($(PsiP.north)+(\horshift)$) -- ($(PhiM.west) -(\vershift)$) node [pos=0.25,anchor=north west,Q node,rotate=45] {$\gen{S}_1,\genr{Q}_1$};

  \draw [arrow,color=OliveGreen,thick] ($(PhiM.east) +(\vershift)$) -- ($(PsiM.north)+(\horshift)$) node [pos=0.25,anchor=south west,Q node,rotate=-45] { $\gen{Q}^2,\genr{S}^2$};
    \draw [arrow,color=Orange,thick] ($(PsiM.north)-(\horshift)$) -- ($(PhiM.east) -(\vershift)$) node [pos=0.25,anchor=north east,Q node,rotate=-45] {$\gen{S}_2,\genr{Q}_2$};

  \draw [arrow,color=OliveGreen,thick] ($(PsiP.south) -(\horshift)$) -- ($(PhiP.west)-(\vershift)$) node [pos=0.75,anchor=north east,Q node,rotate=-45] { $\gen{Q}^2,\genr{S}^2$};
    \draw [arrow,color=Orange,thick] ($(PhiP.west)+(\horshift)$) -- ($(PsiP.south) +(\vershift)$) node [pos=0.75,anchor=south west,Q node,rotate=-45] {$\gen{S}_2,\genr{Q}_2$};

  \draw [arrow,color=OliveGreen,thick] ($(PsiM.south) +(\horshift)$) -- ($(PhiP.east)-(\vershift)$) node [pos=0.75,anchor=north west,Q node,rotate=45] { $\gen{Q}^1,\genr{S}^1$};
    \draw [arrow,color=Orange,thick] ($(PhiP.east)+(\vershift)$) -- ($(PsiM.south) -(\horshift)$) node [pos=0.75,anchor=south east,Q node,rotate=45] {$\gen{S}_1,\genr{Q}_1$};

    \draw [arrow,thick,color=Plum] (PsiM) -- (PsiP) node [pos=0.6,anchor=south west,Q node] {\small $\gen{J}^{AB}_{\bullet}$};
    \draw [arrow] (PsiP) -- (PsiM);
    
  \end{tikzpicture}
\qquad
  \begin{tikzpicture}[%
    box/.style={outer sep=1pt},
    Q node/.style={inner sep=1pt,outer sep=0pt},
    arrow/.style={-latex}
    ]%

    \node [box] (PhiM) at ( 0  , 2.5cm) { $\bar{Z}(p)$};
    \node [box] (PsiP) at (-2.5cm, 0cm) { $\bar{\eta}^1(p)$};
    \node [box] (PsiM) at (+2.5cm, 0cm) { $\bar{\eta}^2(p)$};
    \node [box] (PhiP) at ( 0  ,-2.5cm) { $\bar{Y}(p)$};

    \newcommand{\horshift}{0.09cm,0cm}
    \newcommand{\vershift}{0cm,0.10cm}
 
  \draw [arrow,color=OliveGreen,thick] ($(PhiM.west) +(\vershift)$) -- ($(PsiP.north)-(\horshift)$) node [pos=0.25,anchor=south east,Q node,rotate=45] { $\gen{Q}^1,\genr{S}^1$};
    \draw [arrow,color=Orange,thick] ($(PsiP.north)+(\horshift)$) -- ($(PhiM.west) -(\vershift)$) node [pos=0.25,anchor=north west,Q node,rotate=45] {$\gen{S}_1,\genr{Q}_1$};

  \draw [arrow,color=OliveGreen,thick] ($(PhiM.east) +(\vershift)$) -- ($(PsiM.north)+(\horshift)$) node [pos=0.25,anchor=south west,Q node,rotate=-45] { $\gen{Q}^2,\genr{S}^2$};
    \draw [arrow,color=Orange,thick] ($(PsiM.north)-(\horshift)$) -- ($(PhiM.east) -(\vershift)$) node [pos=0.25,anchor=north east,Q node,rotate=-45] {$\gen{S}_2,\genr{Q}_2$};

  \draw [arrow,color=OliveGreen,thick] ($(PsiP.south) -(\horshift)$) -- ($(PhiP.west)-(\vershift)$) node [pos=0.75,anchor=north east,Q node,rotate=-45] { $\gen{Q}^2,\genr{S}^2$};
    \draw [arrow,color=Orange,thick] ($(PhiP.west)+(\horshift)$) -- ($(PsiP.south) +(\vershift)$) node [pos=0.75,anchor=south west,Q node,rotate=-45] {$\gen{S}_2,\genr{Q}_2$};

  \draw [arrow,color=OliveGreen,thick] ($(PsiM.south) +(\horshift)$) -- ($(PhiP.east)-(\vershift)$) node [pos=0.75,anchor=north west,Q node,rotate=45] { $\gen{Q}^1,\genr{S}^1$};
    \draw [arrow,color=Orange,thick] ($(PhiP.east)+(\vershift)$) -- ($(PsiM.south) -(\horshift)$) node [pos=0.75,anchor=south east,Q node,rotate=45] {$\gen{S}_1,\genr{Q}_1$};

    \draw [arrow,thick,color=Plum] (PsiM) -- (PsiP) node [pos=0.6,anchor=south west,Q node] {\small $\gen{J}^{AB}_{\bullet}$};
    \draw [arrow] (PsiP) -- (PsiM);
    
  \end{tikzpicture}
  \\ 
  \begin{tikzpicture}[%
    box/.style={outer sep=1pt},
    Q node/.style={inner sep=1pt,outer sep=0pt},
    arrow/.style={-latex}
    ]%

    \node [box] (PhiM) at ( 0  , 2.5cm) { $\chi^{\dot{A}}(p)$};
    \node [box] (PsiP) at (-2.5cm, 0cm) { $T^{1\dot{A}}(p)$};
    \node [box] (PsiM) at (+2.5cm, 0cm) { $T^{2\dot{A}}(p)$};
    \node [box] (PhiP) at ( 0  ,-2.5cm) { $\tilde{\chi}^{\dot{A}}(p)$};

    \newcommand{\horshift}{0.09cm,0cm}
    \newcommand{\vershift}{0cm,0.10cm}
 
  \draw [arrow,color=OliveGreen,thick] ($(PhiM.west) +(\vershift)$) -- ($(PsiP.north)-(\horshift)$) node [pos=0.25,anchor=south east,Q node,rotate=45] { $\gen{Q}^1,\genr{S}^1$};
    \draw [arrow,color=Orange,thick] ($(PsiP.north)+(\horshift)$) -- ($(PhiM.west) -(\vershift)$) node [pos=0.25,anchor=north west,Q node,rotate=45] {$\gen{S}_1,\genr{Q}_1$};

  \draw [arrow,color=OliveGreen,thick] ($(PhiM.east) +(\vershift)$) -- ($(PsiM.north)+(\horshift)$) node [pos=0.25,anchor=south west,Q node,rotate=-45] { $\gen{Q}^2,\genr{S}^2$};
    \draw [arrow,color=Orange,thick] ($(PsiM.north)-(\horshift)$) -- ($(PhiM.east) -(\vershift)$) node [pos=0.25,anchor=north east,Q node,rotate=-45] {$\gen{S}_2,\genr{Q}_2$};

  \draw [arrow,color=OliveGreen,thick] ($(PsiP.south) -(\horshift)$) -- ($(PhiP.west)-(\vershift)$) node [pos=0.75,anchor=north east,Q node,rotate=-45] { $\gen{Q}^2,\genr{S}^2$};
    \draw [arrow,color=Orange,thick] ($(PhiP.west)+(\horshift)$) -- ($(PsiP.south) +(\vershift)$) node [pos=0.75,anchor=south west,Q node,rotate=-45] {$\gen{S}_2,\genr{Q}_2$};

  \draw [arrow,color=OliveGreen,thick] ($(PsiM.south) +(\horshift)$) -- ($(PhiP.east)-(\vershift)$) node [pos=0.75,anchor=north west,Q node,rotate=45] { $\gen{Q}^1,\genr{S}^1$};
    \draw [arrow,color=Orange,thick] ($(PhiP.east)+(\vershift)$) -- ($(PsiM.south) -(\horshift)$) node [pos=0.75,anchor=south east,Q node,rotate=45] {$\gen{S}_1,\genr{Q}_1$};

    \draw [arrow,thick,color=Plum] (PsiM) -- (PsiP) node [pos=0.6,anchor=south west,Q node] {\small $\gen{J}^{AB}_{\bullet}$};
    \draw [arrow] (PsiP) -- (PsiM);
    
  \end{tikzpicture}
  \caption{The short representations in which we can accommodate the 8+8 transverse modes of the string. The top two diagrams represent modes related to $AdS_3\times S^3$, while the bottom diagram represents modes related to~$T^4$.
  Note that this picture is valid for $k\neq1$ and $k\neq2$. In particular for $k=1$ we should only consider the $T^4$ modes.}
  \label{fig:representations}
\end{figure}

\begin{table}[t]
\centering
\renewcommand{\arraystretch}{1.5}
 \begin{tabular}{|l|l||c | c| c  | c|} 
 \hline
      & & $\gen{H}$& $\gen{M}$ & $\gen{B}$ & at $k=1$?\\
 \hline
   ferm. & $\chi^{\dot{A}}(p)$ & $\sqrt{\tfrac{k^2}{4\pi^2}p^2+4h^2\sin^2(\tfrac{p}{2})}$ & $\tfrac{k}{2\pi}p$& $\tfrac{k}{2\pi}p-1$&yes\\
 $\text{T}^4$ bos. & $T^{\dot{A}A}(p)$ & $\sqrt{\tfrac{k^2}{4\pi^2}p^2+4h^2\sin^2(\tfrac{p}{2})}$ & $\tfrac{k}{2\pi}p$& $\tfrac{k}{2\pi}p$&yes\\
  ferm. & $\tilde{\chi}^{\dot{A}}(p)$ & $\sqrt{\tfrac{k^2}{4\pi^2}p^2+4h^2\sin^2(\tfrac{p}{2})}$ & $\tfrac{k}{2\pi}p$& $\tfrac{k}{2\pi}p+1$&yes\\
 \hline
 $\text{S}^3$ bos. & $Y(p)$ & $\sqrt{(1+\tfrac{k}{2\pi}p)^2+4h^2\sin^2(\tfrac{p}{2})}$ & $\tfrac{k}{2\pi}p+1$& $\tfrac{k}{2\pi}p-1$&no\\
 ferm. & $\eta^A(p)$ & $\sqrt{(1+\tfrac{k}{2\pi}p)^2+4h^2\sin^2(\tfrac{p}{2})}$ & $\tfrac{k}{2\pi}p+1$& $\tfrac{k}{2\pi}p$&no\\
 $\text{AdS}_3$ bos. & $Z(p)$ & $\sqrt{(1+\tfrac{k}{2\pi}p)^2+4h^2\sin^2(\tfrac{p}{2})}$ & $\tfrac{k}{2\pi}p+1$& $\tfrac{k}{2\pi}p+1$&no\\
 \hline
 $\text{AdS}_3$ bos. & $\bar{Z}(p)$ & $\sqrt{(1-\tfrac{k}{2\pi}p)^2+4h^2\sin^2(\tfrac{p}{2})}$ & $\tfrac{k}{2\pi}p-1$& $\tfrac{k}{2\pi}p-1$&no\\
 ferm. & $\bar{\eta}^A(p)$ & $\sqrt{(1-\tfrac{k}{2\pi}p)^2+4h^2\sin^2(\tfrac{p}{2})}$ & $\tfrac{k}{2\pi}p-1$& $\tfrac{k}{2\pi}p$&no\\
 $\text{S}^3$ bos. & $\bar{Y}(p)$ & $\sqrt{(1-\tfrac{k}{2\pi}p)^2+4h^2\sin^2(\tfrac{p}{2})}$ & $\tfrac{k}{2\pi}p-1$& $\tfrac{k}{2\pi}p+1$&no\\
 \hline
 (``left'' & $Y_\mu(p)$ & $\sqrt{(\mu+\tfrac{k}{2\pi}p)^2+4h^2\sin^2(\tfrac{p}{2})}$ & $\tfrac{k}{2\pi}p+\mu$& $\tfrac{k}{2\pi}p-1$&no\\
 bound & $\eta_\mu^A(p)$ & $\sqrt{(\mu+\tfrac{k}{2\pi}p)^2+4h^2\sin^2(\tfrac{p}{2})}$ & $\tfrac{k}{2\pi}p+\mu$& $\tfrac{k}{2\pi}p$&no\\
 states) & $Z_\mu(p)$ & $\sqrt{(\mu+\tfrac{k}{2\pi}p)^2+4h^2\sin^2(\tfrac{p}{2})}$ & $\tfrac{k}{2\pi}p+\mu$& $\tfrac{k}{2\pi}p+1$&no\\
 \hline
 (``right'' & $\bar{Z}_\mu(p)$ & $\sqrt{(\mu-\tfrac{k}{2\pi}p)^2+4h^2\sin^2(\tfrac{p}{2})}$ & $\tfrac{k}{2\pi}p-\mu$& $\tfrac{k}{2\pi}p-1$&no\\
 bound & $\bar{\eta}_\mu^A(p)$ & $\sqrt{(\mu-\tfrac{k}{2\pi}p)^2+4h^2\sin^2(\tfrac{p}{2})}$ & $\tfrac{k}{2\pi}p-\mu$& $\tfrac{k}{2\pi}p$&no\\
 states) & $\bar{Y}_\mu(p)$ & $\sqrt{(\mu-\tfrac{k}{2\pi}p)^2+4h^2\sin^2(\tfrac{p}{2})}$ & $\tfrac{k}{2\pi}p-\mu$& $\tfrac{k}{2\pi}p+1$&no\\
 \hline

  \hline
 \end{tabular}
 \caption{The physical excitations of the string and their central charges. Each big row is a representation, with the exception of the first big row which describes together both torus representations (for $\dot{A}=1,2$).
 Note that we can only determine the three charges $\gen{H}$, $\gen{M}$, $\gen{B}$ because the fourth charge $\gen{R}$ has been decoupled.
 The details of the allowed representations depend on the value of~$k$. In the pure-RR case, that is at $k=0$, there are eight plus eight ``fundamental particles'' (the transverse modes of the superstring), divided in four representations corresponding to the first three big rows, as well as infinitely many bound-state representations labeled by $\mu=2,3,\dots$.
 For the backgrounds involving NSNS flux, that is $k>0$, we should not consider all values of~$\mu$ if we allow the momentum to take any real value. There are two massless representations corresponding to the torus, plus $(k-1)$ massive representations~\cite{Frolov:2023lwd}. Hence, for $k=1$ there are no massive representations at all, and for $k=2$ the second and third become one and the same representation.}
 \label{tab:representations}
\end{table}

\noindent This, together with the form of the charges~\eqref{eq:centralchargesp} and the consideration that we want $H\geq0$, is sufficient to determine the dispersion relation of the model~\cite{Hoare:2013lja,Lloyd:2014bsa},
\begin{equation}
\label{eq:dispersion}
    H(p) = \sqrt{\left(\mu+\frac{k}{2\pi}p\right)^2+4h^2\sin^2\left(\frac{p}{2}\right)}\,.
\end{equation}
Recall that $\mu\in\mathbb{Z}$ is quantised and it identifies different representations. A perturbative computation is therefore sufficient to determine the allowed values of $\mu$. It turns out that, for generic values of $k$, the transverse excitations of the string arrange themselves into:
\begin{enumerate}
    \item Two irreducible representations with $\mu=0$ corresponding to the four bosons on the torus and to the remaining four fermions. These two representations form a doublet under~$\su(2)_\circ$ (under which the torus fields are charged, see Table~\ref{tab:su2});
    \item One irreducible representation with $\mu=+1$, corresponding to one boson from $\text{AdS}_3$, one boson from the sphere, and two fermions;
    \item One irreducible representation with $\mu=-1$, corresponding to the other boson from $\text{AdS}_3$, the other boson from the sphere, and two more fermions;
    \item Bound-state representations with $\mu=2,3,\dots$ or $\mu=-2,-3,\dots$; however, while for $k=0$ there are infinitely-many such bound states (like in $AdS_5\times S^5$~\cite{Arutyunov:2007tc}), in this case the maximum allowed value of $\mu$ is $k-1$~\cite{Frolov:2023lwd} if the momentum $p$ is allowed to take any real value; we will see that this is the natural choice when we want to compare with the symmetric-product orbifold i.e.~at $k=1$, and comment more on the interpretation of the $\mu\neq0$ representations then. 
\end{enumerate}
A summary of these four irreducible representations is presented in Table~\ref{tab:representations}. The supercharges act on the representations as illustrated by the diagrams in Figure~\ref{fig:representations}.

\subsection{One-particle representations for \texorpdfstring{$k=1$}{k=1}}
Let us now specialise to the $k=1$ case. Here we only need the two irreducible representations corresponding to $\mu=0$, which we can treat in one go thanks to the indices $\dot{A}=1,2$. We have~\cite{Lloyd:2014bsa}
\begin{equation}
\begin{aligned}
    &\big[\gen{Q}^{A},\,\chi^{\dot{A}}(p)\big] = a(p)\,T^{A\dot{A}}(p)\,,\qquad
    &&\big[\gen{Q}^A,\,T^{B\dot{A}}(p)\big] = \varepsilon^{AB}a(p)\,\tilde{\chi}^{\dot{A}}(p)\,,\\
    &\big[\genr{S}^{A},\,\chi^{\dot{A}}(p)\big] = \bar{b}(p)\,T^{A\dot{A}}(p)\,,\qquad
    &&\big[\genr{S}^A,\,T^{B\dot{A}}(p)\big] = \varepsilon^{AB}\bar{b}(p)\,\tilde{\chi}^{\dot{A}}(p)\,,\\
    &\big[\gen{S}^{A},\,T^{B\dot{A}}(p)\big] = \varepsilon^{AB}\bar{a}(p)\,\chi^{\dot{A}}(p)\,,\qquad
    &&\big[\gen{S}^A,\,\tilde{\chi}^{\dot{A}}(p)\big] = \bar{a}(p)\,T^{A\dot{A}}(p)\,,\\
    &\big[\genr{Q}^{A},\,T^{B\dot{A}}(p)\big] = \varepsilon^{AB}b(p)\,\chi^{\dot{A}}(p)\,,\qquad
    &&\big[\genr{Q}^A,\,\tilde{\chi}^{\dot{A}}(p)\big] = b(p)\,T^{A\dot{A}}(p)\,.
\end{aligned}
\end{equation}
This somehow redundant structure is a sign that the representation can be factorised as a tensor product~\cite{Borsato:2013qpa}, much like in $AdS_5\times S^5$; this is convenient for various explicit computations, but not important for us here.
It follows from this algebra that the central charges have eigenvalues
\begin{equation}
    C(p)=a(p)b(p)\,,\qquad C^\dagger(p)=\bar{a}(p)\bar{b}(p)\,,
\end{equation}
while
\begin{equation}
    H(p)=a(p)\bar{a}(p)+b(p)\bar{b}(p),\qquad
    M(p)=a(p)\bar{a}(p)-b(p)\bar{b}(p).
\end{equation}
We  follow the conventions of~\cite{Lloyd:2014bsa} and set
\begin{equation}
\label{eq:reprcoefficients}
\begin{aligned}
    a(p)=e^{+\tfrac{i}{4}p}\sqrt{\frac{ih}{2}(x^-_p -x^+_p)}e^{+i\xi(p)}\,,\qquad
    &&b(p)=-\frac{e^{-\frac{i}{4}p}}{x^-_p}\sqrt{\frac{ih}{2}(x^-_p -x^+_p)}e^{+i\xi(p)}\,,\\
    \bar{a}(p)=e^{-\tfrac{i}{4}p}\sqrt{\frac{ih}{2}(x^-_p -x^+_p)}e^{-i\xi(p)}\,,\qquad
    &&\bar{b}(p)=-\frac{e^{+\frac{i}{4}p}}{x^+_p}\sqrt{\frac{ih}{2}(x^-_p -x^+_p)}e^{-i\xi(p)}\,,
\end{aligned}
\end{equation}
which are written in terms of Zhukovsky variables, whose expression at $k=1$ and $\mu=0$ is 
\begin{equation}
    x^\pm_p=\frac{\tfrac{1}{2\pi}p+\sqrt{\tfrac{1}{4\pi^2}p^2+4h^2\sin^2(p/2)}}{2h\sin^2(p/2)}\,e^{\pm\frac{i}{2}p}\,.
\end{equation}
They satisfy the identities
\begin{equation}
    \frac{x^+_p}{x^-_p}=e^{ip},\quad
    x^+_p+\frac{1}{x^+_p}-x^-_p-\frac{1}{x^-_p}=\frac{2i}{h}M(p),\quad
    x^+_p-\frac{1}{x^+_p}-x^-_p+\frac{1}{x^-_p}=\frac{2i}{h}H(p),
\end{equation}
from which it follows that the central charges have the form which we want.

\subsection{Two-particle representation and ``string-frame'' co-product}
Above we have defined the one-particle representation. A na\"ive way to extend this to a multi-particle representation would be the Leibnitz rule. For instance, for the central charge~$\gen{C}$,
\begin{equation}
    [\gen{C},\,\chi^{A_1}(p_1)\dots\chi^{A_n}(p_n)] = \left(\sum_{i=1}^n C(p_i)\right)\,\chi^{A_1}(p_1)\dots\chi^{A_n}(p_n),\  \text{(trivial coproduct).}
\end{equation}
However, this is incompatible with the dependence of the central charges on the momentum~\eqref{eq:centralchargesp}. In particular we can check that
\begin{equation}
    C(p_1+p_2)\neq C(p_1)+C(p_2)\,,
\end{equation}
in general. In fact, these considerations hold regardless of the value of the function~$\xi(p)$.

We therefore need to introduce a non-trivial co-product. This can be done by keeping track of $\xi$-dependence of multi-particle representations~\cite{Arutyunov:2006yd} or by constructing an ad-hoc coproduct~\cite{Lloyd:2014bsa}. We obtain
\begin{equation}
\label{eq:coproduct-string}
\begin{aligned}
    Q^A(p_1,p_2)&=Q^A(p_1)\otimes 1\,e^{+i\xi(p_2)}+e^{+\tfrac{i}{2}p_1+i\xi(p_1)}\,\Sigma\otimes Q^{A}(p_2)\,,\\
    \widetilde{Q}^A(p_1,p_2)&=\widetilde{Q}^A(p_1)\otimes 1\,e^{+i\xi(p_2)}+e^{+\tfrac{i}{2}p_1+i\xi(p_1)}\,\Sigma\otimes \widetilde{Q}^{A}(p_2)\,,\\
    S^A(p_1,p_2)&=S^A(p_1)\otimes 1\,e^{-i\xi(p_2)}+e^{-\tfrac{i}{2}p_1-i\xi(p_1)}\,\Sigma\otimes S^{A}(p_2)\,,\\
    \widetilde{S}^A(p_1,p_2)&=\widetilde{S}^A(p_1)\otimes 1\,e^{-i\xi(p_2)}+e^{-\tfrac{i}{2}p_1-i\xi(p_1)}\,\Sigma\otimes \widetilde{S}^{A}(p_2)\,,
\end{aligned}
\end{equation}
where $\Sigma=(-1)^F$ is the fermion-sign matrix, which we made explicit as we are dealing with supercharges, and the one-particle representation is that discussed in the previous section. For the central charges, this gives
\begin{equation}
    C(p_1,p_2)=C(p_1)+e^{ip_1}C(p_2)=C(p_1+p_2)\,,
\end{equation}
and similarly for $C^\dagger$,
while the co-product of $H$ and $M$ remains unchanged
\begin{equation}
    H(p_1,p_2)=H(p_1)+H(p_2)\,,\qquad
    M(p_1,p_2)=M(p_1)+M(p_2)\,.
\end{equation}
This way of defining the co-product is natural in string theory, and is especially convenient when considering the mirror model. We will see however that there is another co-product prescription, which is sometimes called ``the spin-chain frame''.

\subsection{Two-particle S matrix}
Having fixed the two-particle representation, we can impose that the two-particle S-matrix $\mathcal{S}(p_1,p_2)$ commutes with all the linear constraints
\begin{equation}
\begin{aligned}
    &Q^A(p_2,p_1)\, \mathcal{S}(p_1,p_2)=\mathcal{S}(p_1,p_2)\,Q^A(p_1,p_2)\,,\\
    &\widetilde{Q}^A(p_2,p_1)\, \mathcal{S}(p_1,p_2)=\mathcal{S}(p_1,p_2)\,\widetilde{Q}^A(p_1,p_2)\,,\\
    &S^A(p_2,p_1)\, \mathcal{S}(p_1,p_2)=\mathcal{S}(p_1,p_2)\,S^A(p_1,p_2)\,,\\
    &\widetilde{S}^A(p_2,p_1)\, \mathcal{S}(p_1,p_2)=\mathcal{S}(p_1,p_2)\,\widetilde{S}^A(p_1,p_2)\,,
\end{aligned}
\end{equation}
where we can represent $\mathcal{S}(p_1,p_2)$ explictly as a matrix.
It is known that, for each choice of irreducible representations, this determines $\mathcal{S}(p_1,p_2)$ uniquely up to an overall normalisation, the so-called dressing factor~\cite{Lloyd:2014bsa}. Moreover, the resulting S-matrix automatically satisfies the Yang-Baxter equation.
The explicit form of $\mathcal{S}(p_1,p_2)$ may be found in~\cite{Lloyd:2014bsa}.

\section{The symmetric-product orbifold limit}
In the previous section we recapped the results of~\cite{Lloyd:2014bsa}, valid for any $k, h$, and specialised them to the case of $k=1$ and $h$ arbitrary. Now we want to further specialise them to the case where $h\ll1$, \textit{i.e.}\ around the symmetric-product CFT of~$T^4$.

\begin{figure}[t]
\centering
\begin{tikzpicture}
  \begin{axis}[
    clip = true,
    clip mode=individual,
    restrict y to domain=0:8,
    axis x line = middle,
    axis y line = middle,
    domain = -5:5,
    xmin = -5,
    xmax = 5,
    xtick={-4,-3,-2,-1,0,1,2,3,4},
    xticklabels={$-4$,$-3$,$-2$,$-1$,$0$,$1$,$2$,$3$,$4$},
    ytick={2,4,6},
    yticklabels={$2$,$4$,$6$},
    ymin = 0,
    ymax = 8,
    after end axis/.code={\path (axis cs:0,0) node [anchor=north west,yshift=-0.075cm,xshift=-0.075cm] {0} 
    ;},
  ]

    \addplot[color=Cyan,samples=200,smooth,ultra thick] {1/2*(sqrt((x)*(x)+16*sin(180*x)*sin(180*x))+x)} node[above,pos=1] {};
    \addplot[color=Black,samples=200,smooth,ultra thick,dashed] {1/2*(sqrt((x)*(x)+0.16*sin(180*x)*sin(180*x))+x)} node[below,pos=2] {};

  \end{axis}
  \node at (2.6,5.5) {$\gen{L}_0-\gen{J}^3$};
  \node at (6.5,0.5) {$\displaystyle\frac{p}{2\pi}$};
\end{tikzpicture}%
\hspace*{0.6cm}
\begin{tikzpicture}
  \begin{axis}[
    clip = true,
    clip mode=individual,
    restrict y to domain=0:8,
    axis x line = middle,
    axis y line = middle,
    domain = -5:5,
    xmin = -5,
    xmax = 5,
    xtick={-4,-3,-2,-1,0,1,2,3,4},
    xticklabels={$-4$,$-3$,$-2$,$-1$,$0$,$1$,$2$,$3$,$4$},
    ytick={2,4,6},
    yticklabels={$2$,$4$,$6$},
    ymin = 0,
    ymax = 8,
    after end axis/.code={\path (axis cs:0,0) node [anchor=north west,yshift=-0.075cm,xshift=-0.075cm] {0} 
    ;},
  ]

    \addplot[color=Cyan,samples=200,smooth,ultra thick] {1/2*(sqrt((x)*(x)+16*sin(180*x)*sin(180*x))-x)} node[above,pos=1] {};
    \addplot[color=Black,samples=200,smooth,ultra thick,dashed] {1/2*(sqrt((x)*(x)+0.16*sin(180*x)*sin(180*x))-x)} node[below,pos=2] {};

  \end{axis}
  \node at (4.3,5.5) {$\genr{L}_0-\genr{J}^3$};
  \node at (0.5,0.5) {$\displaystyle\frac{p}{2\pi}$};
\end{tikzpicture}
\caption{We plot the left and right part of the energy, given by $\tfrac{1}{2}(\gen{H}\pm\gen{M})$, for two excitations in the $\mu=0$ representation. The solid cyan line corresponds to $k=1$ and $h=2$; the dashed blue line corresponds to $k=1$ and $h=0.2$.
We see that  for any $h\geq0$, the left part of the Hamiltonian vanishes at $p=-2\pi, -4\pi, \dots$, and the right part vanishes for $p=0,2\pi,\dots$ (that is, for physical states).
When $h\to0$, each plot degenerates into a piecewise-linear function of the form $\pm\frac{1}{2\pi}p\,\Theta(\pm p)$, where $\Theta(p)$ is the Heaviside function.}
\label{fig:dispersionRL}
\end{figure}

\subsection{Free symmetric orbifold dispersion at \texorpdfstring{$h=0$}{h=0}}
We expect that the unperturbed symmetric-product CFT of~$T^4$ should arise from the $h\to0$ limit of the formulae of the above section. Indeed, this is the setup where the background is supported \textit{only} by the Kalb-Ramond $B$-field and by the $H=dB$ flux. In this case we have that
the shortening condition~\eqref{eq:shortening} becomes
\begin{equation}
    H^2=M^2\,,\qquad
    \text{that is}\qquad
    \big(L_0-J^3\big)\big(\tilde{L}_0-\tilde{J}^3\big)=0\,.
\end{equation}
In other words, the excitations may be charged under the chiral part of the lightcone Hamiltonian, or under the anti-chiral part, but not under both. More explicitly we have, for any $k$
\begin{equation}
\label{eq:chiraldispersion}
H= \left|\mu+\frac{k}{2\pi}p\right|\,,\qquad
M=\mu+\frac{k}{2\pi}p\,,\qquad|\mu|\leq k-1\,,
\end{equation}
which for the special case of $k=1$ is simply
\begin{equation}
\label{eq:chiraldispersionkis1}
H= \frac{1}{2\pi}|p|\,,\qquad
M=\frac{1}{2\pi}p\,,
\end{equation}
which means that
\begin{equation}
    L_0-J^3 = \frac{p}{2\pi}\,\Theta(p)\,,\qquad
    \tilde{L}_0-\tilde{J}^3 = -\frac{p}{2\pi}\,\Theta(-p)\,,
\end{equation}
where $\Theta$ is the Heaviside step function, see also Figure~\ref{fig:dispersionRL}.

At generic~$k$, this chiral structure matches perfectly with the worldsheet description of the $\mathfrak{sl}(2)\oplus\su(2)$ Wess-Zumino-Witten model~\cite{Dei:2018mfl}. In the specific case of $k=1$, it also matches with the description of the states in the symmetric-product orbifold of $\text{T}^4$. Indeed the symmetric-product orbifold CFT is naturally split into chiral and anti-chiral excitations. This was an early argument to propose that the symmetric-orbifold of $\text{T}^4$ should be dual to a pure-NSNS worldsheet model~\cite{Sfondrini:talk}, and more specifically to the model at $k=1$, as it was subsequently established by CFT techniques~\cite{Eberhardt:2018ouy}.

These excitations can be easily matched with those appearing in the symmetric-product orbifold of~$T^4$, $\text{Sym}_N(T^4)$, in the $N\to\infty$ limit (corresponding to the free string theory). We identify the $\psu(1,1|2)^{\oplus2}$ with the globally defined generators of $\mathcal{N}=(4,4)$ superconformal symmetry. In particular $\gen{L}_0$ is the chiral scaling generator, and $\gen{J}^3$ is the chiral Cartan generator of R-symmetry. Consider a sector where $w$ copies of the symmetric product orbifold form a $w$-cycle, and the remaining $N-w$ copies are trivial 1-cycles. Such a state can be created by acting on the CFT vacuum with a $w$-twist field~$\sigma_w$.
Due to the presence of the twist field, the ``ground state'' of this twisted sector is not BPS: its energy under $\gen{L}_0+\genr{L}_0$ depends on $w$, and it is greater than zero;%
\footnote{Specifically, for $w$ odd it is $(w-1/w)/4$ and for $w$ even it is~$w/4$; this slight difference is due to the contribution of the fermion zero-modes.}
its  $\mathfrak{sl}(2)$ spin $\gen{L}_0-\genr{L}_0$ is zero, and so are its R-symmetry charges. 
In the vicinity of the twist field~$\sigma_w$, the fundamental modes of the bosons and fermion fields are fractionally-moded, and they take the form
\begin{equation}
\label{eq:CFTexctiations}
    \alpha^{A\dot{A}}_{-\frac{n}{w}},\quad
    \tilde{\alpha}^{A\dot{A}}_{-\frac{n}{w}},\quad
    \Psi^{\alpha \dot{A}}_{-\frac{1}{2}-\frac{n}{w}},\quad
    \tilde{\Psi}^{\dot{\alpha} \dot{A}}_{-\frac{1}{2}-\frac{n}{w}}\,.
\end{equation}
Because some of the fermion modes have negative lightcone energy $\gen{L}_0-\gen{J}^3$, it is possible to construct several BPS states from the twisted vacuum~\cite{Lunin:2000yv} by filling  a Fermi sea with $\Psi^{\alpha \dot{A}}_{-\frac{1}{2}-\frac{n}{w}}$'s and $\tilde{\Psi}^{\dot{\alpha} \dot{A}}_{-\frac{1}{2}-\frac{n}{w}}$'s (with their $n\geq0$).
In particular, there is one (and only one) state~$|0\rangle_w$ with
\begin{equation}
    \gen{L}_0|0\rangle_w=\genr{L}_0|0\rangle_w=
    \gen{J}^3|0\rangle_w=\genr{J}^3|0\rangle_w=\frac{w}{2}|0\rangle_w\,,
\end{equation}
and without any charge under~$\su(2)_{\circ}$.
Up to a normalisation, this is schematically%
\footnote{Strictly speaking, this expression is valid for $w$ odd; for $w$ even one has to be mindful of the existence of a Clifford module worth of states, so that the twist field can be written as $\sigma_{2m}^{\alpha\dot\alpha}$; in this formula, we are implicitly choosing $\sigma_{2m}^{--}$ as a reference state and acting to get the desired R-charge~\cite{Lunin:2000yv}.}
\begin{equation}
    |0\rangle_w=\varepsilon_{\dot{A}\dot{B}}\Psi^{+ \dot{A}}_{-\frac{1}{2}}\tilde{\Psi}^{+ \dot{B}}_{-\frac{1}{2}}\prod_{n=1}^{\lfloor\frac{w}{2}\rfloor}\left(\Psi^{+ \dot{1}}_{-\frac{1}{2}+\frac{n}{w}}\Psi^{+ \dot{2}}_{-\frac{1}{2}+\frac{n}{w}}\tilde{\Psi}^{+ \dot{1}}_{-\frac{1}{2}+\frac{n}{w}}\tilde{\Psi}^{+ \dot{2}}_{-\frac{1}{2}+\frac{n}{w}}\right)\sigma_w|0\rangle\,,
\end{equation}
where $|0\rangle$ is the CFT vacuum.
Since this state is BPS, it is annihilated by the supercharges $\gen{Q}^A$, $\genr{Q}^A$, $\gen{S}^A$, $\genr{S}^A$ introduced in the previous section. As a result, all the excitations on $|0\rangle_w$ must transform in a representation of the symmetry algebra generated by $\gen{Q}^A$, $\genr{Q}^A$, $\gen{S}^A$ and $\genr{S}^A$. It is clear that the excitations which do not annihilate the vacuum are those in~\eqref{eq:CFTexctiations} with $n>0$ (or $n=0$ in the case of the fermion zero-modes).
Furthermore, we immediately see that
\begin{equation}
\begin{aligned}
 \relax    [\gen{L}_0-\gen{J}^3,\alpha^{A\dot{A}}_{-\frac{n}{w}}]&=\frac{n}{w}\,\alpha^{A\dot{A}}_{-\frac{n}{w}},
    \qquad
    &[\genr{L}_0-\genr{J}^3,\tilde\alpha^{A\dot{A}}_{-\frac{n}{w}}]&=\frac{n}{w}\,\tilde\alpha^{A\dot{A}}_{-\frac{n}{w}},\\
    [\gen{L}_0-\gen{J}^3,\Psi^{+\dot{A}}_{-\frac{1}{2}-\frac{n}{w}}]&=\frac{n}{w}\,\Psi^{+\dot{A}}_{-\frac{1}{2}-\frac{n}{w}},\qquad
    &[\genr{L}_0-\genr{J}^3,\tilde\Psi^{+\dot{A}}_{-\frac{1}{2}-\frac{n}{w}}]&=\frac{n}{w}\,\tilde\Psi^{+\dot{A}}_{-\frac{1}{2}-\frac{n}{w}},
    \\
    [\gen{L}_0-\gen{J}^3,\Psi^{-\dot{A}}_{+\frac{1}{2}-\frac{n}{w}}]&=\frac{n}{w}\,\Psi^{-\dot{A}}_{+\frac{1}{2}-\frac{n}{w}},\qquad
    &[\genr{L}_0-\genr{J}^3,\tilde\Psi^{-\dot{A}}_{+\frac{1}{2}-\frac{n}{w}}]&=\frac{n}{w}\,\tilde\Psi^{-\dot{A}}_{+\frac{1}{2}-\frac{n}{w}},
\end{aligned}
\end{equation}
while the chiral generators annihilate the anti-chiral excitations, and viceversa.
The remaining charges can be easily found, and they are collected in table~\ref{tab:cftcartan}.

\begin{table}[t]
    \centering

\begin{center}
\renewcommand{\arraystretch}{1.5}
\begin{tabular}{|c|cc|cc|cc||ccc|c|}
\hline
Mode  & $\gen{L}_0$ & $\genr{L}_0$ & $-\gen{J}^3$ &$-\genr{J}^3$&$\mathfrak{su}(2)_{\bullet}$&$\mathfrak{su}(2)_{\circ}$& $\gen{H}$ & $\gen{M}$ & $\gen{B}$&State\\
\hline
$\Psi^{-\dot{A}}_{+\tfrac{1}{2}-\tfrac{n}{w}}$ & $-\frac{1}{2}+\frac{n}{w}$ & 0 & $+\frac{1}{2}$ & 0 & \textbf{1}& \textbf{2}&$\frac{n}{w}$&$\frac{n}{w}$&$\frac{n}{w}-1$&$\chi^{\dot{A}}(p)$\\
$\alpha^{A\dot{A}}_{-\tfrac{n}{w}}$ & $~0+\frac{n}{w}$ & 0 & $~0$ & 0 & \textbf{2}& \textbf{2}&$\frac{n}{w}$&$\frac{n}{w}$&$+\frac{n}{w}$&$T^{A\dot{A}}(p)$\\
$\Psi^{+\dot{A}}_{-\tfrac{1}{2}-\tfrac{n}{w}}$ & $+\frac{1}{2}+\frac{n}{w}$ & 0 & $-\frac{1}{2}$ & 0 & \textbf{1}& \textbf{2}&$\frac{n}{w}$&$\frac{n}{w}$&$\frac{n}{w}+1$&$\tilde\chi^{\dot{A}}(p)$\\
\hline
$\tilde{\Psi}^{+\dot{A}}_{-\tfrac{1}{2}-\tfrac{n}{w}}$ & 0 & $+\frac{1}{2}+\frac{n}{w}$  & 0  & $-\frac{1}{2}$& \textbf{1}& \textbf{2}&$\frac{n}{w}$&$-\frac{n}{w}$&$-\frac{n}{w}-1$&$\chi^{\dot{A}}(-p)$\\
$\tilde{\alpha}^{A\dot{A}}_{-\tfrac{n}{w}}$ & 0 &$~0+\frac{n}{w}$ & 0 & 0 & \textbf{2}& \textbf{2}&$\frac{n}{w}$&$-\frac{n}{w}$&$-\frac{n}{w}$&$T^{A\dot{A}}(-p)$\\
$\tilde{\Psi}^{-\dot{A}}_{+\tfrac{1}{2}-\tfrac{n}{w}}$  & 0 & $-\frac{1}{2}+\frac{n}{w}$ &  0 & $+\frac{1}{2}$ & \textbf{1}& \textbf{2}&$\frac{n}{w}$&$-\frac{n}{w}$&$-\frac{n}{w}+1$&$\tilde\chi^{\dot{A}}(-p)$\\
\hline
$\gen{Q}^A$&$+\frac{1}{2}$&0&$-\frac{1}{2}$&0&\textbf{2}&\textbf{1}&0&0&$+1$&\\
$\gen{S}^A$&$-\frac{1}{2}$&0&$+\frac{1}{2}$&0&\textbf{2}&\textbf{1}&0&0&$-1$&\\
$\genr{Q}^A$&0&$+\frac{1}{2}$&0&$-\frac{1}{2}$&\textbf{2}&\textbf{1}&0&0&$-1$&\\
$\genr{S}^A$&0&$-\frac{1}{2}$&0&$+\frac{1}{2}$&\textbf{2}&\textbf{1}&0&0&$+1$&\\
\hline
\end{tabular}
\end{center}
    \caption{Cartan charges of the excitations over the BPS vacuum of the Sym$_N T^4$ theory and their corresponding state in terms of the representations of Figure~\ref{fig:representations}, with the identification $p=2\pi n/w$ in the convention $p>0$. We also list the Cartan charges of the supercharges which annihilate $|0\rangle_w$; they cannot be interpreted in terms of single excitations with $\mu=0$, as we will discuss below.}
    \label{tab:cftcartan}
\end{table}

Comparing table~\ref{tab:cftcartan} with the values in table~\ref{tab:representations} at $h=0$ immediately suggests the identifications%
\footnote{One should not confuse the tilde on $\tilde{\chi}$ with the one arising from chiral/antichiral fields in the CFT, except when studying the zero-modes of the fermions.}
\begin{equation}
\label{eq:chiralident}
    \alpha^{A\dot{A}}_{-\frac{n}{w}} = T^{A\dot{A}}(p)\,,\qquad
    \Psi^{+\dot{A}}_{-\frac{1}{2}-\frac{n}{w}} = \tilde{\chi}^{\dot{A}}(p),\qquad
    \Psi^{-\dot{A}}_{+\frac{1}{2}-\frac{n}{w}} = {\chi}^{\dot{A}}(p),\qquad p=\frac{2\pi n}{w}\geq 0,
\end{equation}
as well as
\begin{equation}
\label{eq:antichiralident}
    \tilde\alpha^{A\dot{A}}_{-\frac{n}{w}} = T^{A\dot{A}}(p)\,,\qquad
    \tilde\Psi^{+\dot{A}}_{-\frac{1}{2}-\frac{n}{w}} = {\chi}^{\dot{A}}(p),\qquad
    \tilde\Psi^{-\dot{A}}_{+\frac{1}{2}-\frac{n}{w}} = \tilde{\chi}^{\dot{A}}(p),\qquad p=-\frac{2\pi n}{w}\leq 0.
\end{equation}
One should be slightly careful in keeping track of fermion zero-modes, as explained in~\cite{Eden:2021xhe}.
Finally, note that with this identification the level-matching condition~\eqref{eq:levelmatching} turns into the orbifold-invariance condition, \textit{i.e.}\ that the sum of the chiral and anti-chiral fractionary modes should be integer.

Having discussed the representations at any~$h$ in~\eqref{eq:reprcoefficients}, it is easy to take the $h\to0$ limit and find
\begin{equation}
\label{eq:reprcoefficientshiszero}
\begin{aligned}
    a(p)=e^{+\tfrac{i}{4}p+i\xi(p)}\sqrt{\frac{p}{2\pi}}\Theta(+p)\,,\qquad
    &&b(p)=s\,e^{+\tfrac{i}{4}p+i\xi(p)}\sqrt{\frac{-p}{2\pi}}\Theta(-p)\,,\\
    \bar{a}(p)=e^{-\tfrac{i}{4}p-i\xi(p)}\sqrt{\frac{p}{2\pi}}\Theta(+p)\,,\qquad
    &&\bar{b}(p)=s\,e^{-\tfrac{i}{4}p-i\xi(p)}\sqrt{\frac{-p}{2\pi}}\Theta(-p)\,,
\end{aligned}
\end{equation}
where $s=-\text{sign}[\sin(p/2)]$ can be reabsorbed for instance by redefining  the fermions \textit{e.g.}\ $\chi(p)\to s\chi(p)$ and $\tilde\chi(p)\to s\tilde\chi(p)$. 
In any case, it is clear that in this limit the representation is split into two parts,
\begin{equation}
\label{eq:chiral-antichiral}
    p>0\ \text{which we call ``chiral''},\qquad
    p<0\ \text{which we call ``antichiral''},
\end{equation}
for $\mu=0$ modes.

\subsection{On the representations with \texorpdfstring{$\mu\neq0$}{mu!=0}}
\label{sec:orbifold:boundstates}
We have discussed in and around Table~\ref{tab:representations} that, for $k=1$, we may restrict to representations with~$\mu=0$ as long as we consider excitations with arbitrary momentum~$p\in\mathbb{R}$.
Let us see why this is the case. First, we note that even when $h>0$, in the $w=\infty$ theory, the particle content of the theory as well as the form of the S~matrix is determined by representation theory in terms of the allowed ``asymptotic states'', which in turn are build out of asymptotic particle excitations of definite
momentum $p$.
Short irreducible representations with the same values of the central charges are isomorphic and yield indistinguishable particles.
Note that the eigenvalues of the central charges $\gen{H}$, $\gen{M}$, $\gen{B}$, which determine the representation coefficients, are unchanged under the simultaneous shift
\begin{equation}
\label{eq:kperiodicity}
    \mu\to\mu\pm k\,,\qquad p\to p\mp 2\pi\,.
\end{equation}
This means that for $k=1$ all representations can be obtained by taking only $\mu=0$ representations and allowing (the real part of) $p$ to take
any value. This was also discussed in~\cite{Frolov:2023lwd}.

\begin{table}[t]
    \centering

\begin{center}
\renewcommand{\arraystretch}{1.5}
\begin{tabular}{|c|cc|cc|c||ccc|c|}
\hline
Mode  & $\gen{L}_0$ & $\genr{L}_0$ & $-\gen{J}^3$ &$-\genr{J}^3$&$\mathfrak{su}(2)_{\bullet}$& $\gen{H}$ & $\gen{M}$ & $\gen{B}$&State\\
\hline
$\gen{J}^{--}_{-\tfrac{n}{w}}$ & $0+\frac{n}{w}$ & 0 & $+1$ & 0 & \textbf{1}&$1+\frac{n}{w}$&$1+\frac{n}{w}$&$\frac{n}{w}-1$&$Y(p)$\\
$\gen{Q}^{-A}_{-\frac{1}{2}-\tfrac{n}{w}}$ & $\frac{1}{2}+\frac{n}{w}$ & 0 & $+\tfrac{1}{2}$ & 0 & \textbf{2}& $1+\frac{n}{w}$&$1+\frac{n}{w}$&$\frac{n}{w}$&$\eta^{A}(p)$\\
$\gen{L}_{-1-\tfrac{n}{w}}^{\text{(low)}}$ & $1+\frac{n}{w}$ & 0 & $~0$ & 0 & \textbf{1}&$1+\frac{n}{w}$&$1+\frac{n}{w}$&$\frac{n}{w}+1$&$Z(p)$\\
\hline
$\genr{L}^{\text{(high)}}_{-1-\tfrac{n}{w}}$& 0  & $1+\frac{n}{w}$ & 0 & $~0$ & \textbf{1}&$1+\frac{n}{w}$&$-1-\frac{n}{w}$&$-\frac{n}{w}-1$&$\bar{Z}(-p)$\\
$\genr{Q}^{-A}_{-\frac{1}{2}-\tfrac{n}{w}}$& 0 & $\frac{1}{2}+\frac{n}{w}$ & 0 & $+\tfrac{1}{2}$ & \textbf{2}& $1+\frac{n}{w}$&$-1-\frac{n}{w}$&$-\frac{n}{w}$&$\bar{\eta}^{A}(-p)$\\
$\genr{J}^{--}_{-\tfrac{n}{w}}$& 0 & $0+\frac{n}{w}$  & 0 & $+1$  & \textbf{1}&$1+\frac{n}{w}$&$-1-\frac{n}{w}$&$-\frac{n}{w}+1$&$\bar{Y}(-p)$\\
\hline
$\gen{L}^{\text{(high)}}_{+1-\tfrac{n}{w}}$& $-1+\frac{n}{w}$ & 0 & $~0$ & 0  & \textbf{1}&$-1+\frac{n}{w}$&$-1+\frac{n}{w}$&$\frac{n}{w}-1$&$\bar{Z}(p)$\\
$\gen{Q}^{+A}_{+\frac{1}{2}-\tfrac{n}{w}}$ & $-\frac{1}{2}+\frac{n}{w}$ & 0 & $-\tfrac{1}{2}$ & 0  & \textbf{2}& $-1+\frac{n}{w}$&$-1+\frac{n}{w}$&$\frac{n}{w}$&$\bar{\eta}^{A}(p)$\\
$\gen{J}^{++}_{-\tfrac{n}{w}}$ & $0+\frac{n}{w}$  & 0  & $-1$ & 0  & \textbf{1}&$-1+\frac{n}{w}$&$-1+\frac{n}{w}$&$\frac{n}{w}+1$&$\bar{Y}(p)$\\
\hline
$\genr{J}^{++}_{-\tfrac{n}{w}}$ & 0 & $+\frac{n}{w}$ & 0 & $-1$ & \textbf{1}&$-1+\frac{n}{w}$&$1-\frac{n}{w}$&$-\frac{n}{w}-1$&$Y(-p)$\\
$\genr{Q}^{+A}_{+\frac{1}{2}-\tfrac{n}{w}}$ & 0 & $-\frac{1}{2}+\frac{n}{w}$& 0 & $-\tfrac{1}{2}$  & \textbf{2}& $-1+\frac{n}{w}$&$1-\frac{n}{w}$&$-\frac{n}{w}$&$\eta^{A}(-p)$\\
$\genr{L}_{+1-\tfrac{n}{w}}^{\text{(low)}}$& 0 & $-1+\frac{n}{w}$  &0 & $~0$  & \textbf{1}&$-1+\frac{n}{w}$&$1-\frac{n}{w}$&$-\frac{n}{w}+1$&$Z(-p)$\\
\hline
\end{tabular}
\end{center}
    \caption{Cartan charges of modes of the currents. The charge under $\gen{B}$ can be used to identify the ``slope'' of the momentum, i.e.~if the excitations should be identified with particles with group velocity $\tfrac{\partial}{\partial p} H(p)  =+\tfrac{1}{2\pi}$ (chiral) or $\tfrac{\partial}{\partial p} H(p) =-\tfrac{1}{2\pi}$ (antichiral). In terms of the momentum, chiral particles have $p>-2\pi\mu$, generalising~\eqref{eq:chiral-antichiral}. The first block of three excitations corresponds to chiral excitations with~$\mu=+1$, while the second block corresponds to antichiral particles with $\mu=-1$. The generator $\gen{L}^{\text{(low)}}_{-n/w}$ is actually a linear combination of $\gen{L}$ and $\gen{J}^3$ such that $[\gen{Q}^A,\gen{L}^{\text{(low)}}_{-n/w}]=0$. Similarly, $\gen{L}^{\text{(high)}}_{-n/w}$ is given by the orthogonal combination, so that $[\gen{S}^A,\gen{L}^{\text{(low)}}_{-n/w}]=0$.
    The first two blocks contain the lowering operators of $\mathfrak{psu}(1,1|2)_L\oplus\mathfrak{psu}(1,1|2)_R$ (cf.~Table~\ref{tab:cftcartan}) which arise at $n=p=0$.
    The remaining two blocks, which we include for completeness, contain the other two ``slopes'' of the dispersion relation: antichiral excitations with $\mu=+1$ and chiral ones with $\mu=-1$; they only exist for $n>2w$ (correspondingly, for $\mu p<-2\pi$).}
    \label{tab:munotzero}
\end{table}

This, and a principle of economy, makes us conjecture that the $\mu=0$ representations give the whole spectrum of the theory.%
\footnote{Note that from the point of view of counting representations up to isomorphisms, this is equivalent to allowing $\mu$ to take integer values, but restricting $p$ to a fundamental region of size $2\pi$.}
Of course in principle it could be that isomorphic representations should be counted multiple times --- for instance, that the spectrum of asymptotic states includes particles of any real $p$ with several values of $\mu$, which
due to~\eqref{eq:kperiodicity} results in degeneracies. To rule this out we note that the
partition function of the undeformed orbifold theory is obtained from the $\mu=0$ excitations alone, and that the partition function of the deformed theory is to be regular for small $h$ (if a bunch of new states appeared at $\mathcal{O}(h)$, one could not do conformal perturbation theory). 

To clarify this let us identify the $\mu=\pm1$ representations in terms symmetric-product orbifold fields. We expect those modes to be related to the generators of $\mathfrak{psu}(1,1|2)_L\oplus\mathfrak{psu}(1,1|2)_R$~\cite{Gomis:2002qi,Gava:2002xb}. Indeed, the fractionary modes of the currents fit precisely in the ``diamond'' structure of the modules of Figure~\ref{fig:representations} with the desired charges of $\gen{H},\gen{M},\gen{B}$.%
\footnote{A.S.~would like to thank Olof Ohlsson Sax for many discussions on this point, leading to the construction of Table~\ref{tab:munotzero} in spring 2015 at the GGI in Florence.}
It is worth noting that the charge under~$\gen{R}$ \textit{is not the same} as for the representations Tables~\ref{tab:charges} and~\ref{tab:munotzero}, but of course the charge $\gen{R}$ decouples in the decompactification limit~$w\to\infty$.
It is then clear that $\mu=\pm1$ modes and $\mu=0$ modes are not independent. For instance%
\footnote{
This relation is valid at $h=0$ and gets corrected order by order in~$h$.}
\begin{equation}
\label{eq:compositeJ}
    J^{++}_{-\tfrac{n}{w}} = 
    \sum_{\ell} \varepsilon_{\dot{A}\dot{B}}\Psi^{+\dot{A}}_{-\frac{1}{2}-\tfrac{n-\ell}{w}}\Psi^{+\dot{B}}_{+\frac{1}{2}-\tfrac{\ell}{w}}\,,
\end{equation}
which indicates how the $\mu=\pm1$ excitations are bilinear combinations of fundamental modes. Once again, counting \textit{all} modes with $\mu=0$ as well as all $|\mu|=1$ modes would be overcounting; the same goes with representations with $\mu=\pm2,\pm3, \dots$ which should be identified with multilinear combinations of the fundamental fields.
We rather should restrict to $\mu=0$ and consider any value of the real part of the momentum.

Finally we note that in~\eqref{eq:compositeJ}, \textit{i.e.}~at $h=0$, the currents should not be considered as \textit{bound states}.
While there is a similarity with the bound states of the pure-RR model (see \textit{e.g.}~\cite{Seibold:2022mgg}) in that the  R-charges add up as expected, at $h=0$ the dispersion relation is linear and there is no binding energy. Hence, we should consider these as  \textit{composite states}, rather than bound states, at $h=0$.
We will return later to the issue of bound states for~$h>0$.

\subsection{Small-\texorpdfstring{$h$}{h} perturbation and length-changing effects}
\label{sec:orbifold:lenghtchanging}

In a similar way, we can find the prediction from integrability at the next subleading order in $h$, which amounts to turning on a bit of RR flux.
In terms of the symmetric-product orbifold, this is related to deforming the theory with a marginal operator from the $w=2$ sector of the theory. Namely, up to a normalisation the deforming operator corresponds to the state
\begin{equation}\label{eq:deformation}
    |\mathcal{O}\rangle=\varepsilon_{\alpha\beta}\varepsilon_{\dot\alpha\dot\beta}\varepsilon_{AB}\gen{Q}^{\alpha A}\genr{Q}^{\dot{\alpha}B}\sigma_{2}^{\beta\dot\beta}|0\rangle\,,
\end{equation}
where the indices on $\sigma_2$ indicate the states in the Clifford module of the twist-2 sector. In conformal perturbation theory, one can then find the corrections to various observables by repeated insertions of~$\mathcal{O}$. The resulting correlation functions should obey a number of selection rules due to orbifold invariance (\textit{ie.}\ due to the fact that they must be single-valued). For instance, a three point function involving an operator on a $w$-cycle and $\mathcal{O}$ must necessarily involve an operator on a $(w\pm1)$ cycle in order to be non-vanishing.

We can describe the expected outcome of such a computation from our finite-$h$ representations.
As we can see already from the finite-$h$ dispersion, as well as pictorially from figure~\ref{fig:dispersionRL}, we expect that at this order chiral excitation will not be annihilated by the anti-chiral generators, and vice-versa.
For instance, let us consider a chiral excitation, for instance $T^{A\dot{A}}(p)$ with $p=2\pi n/w>0$, which by our map~\eqref{eq:chiraldispersion} must correspond to $\alpha^{A\dot{A}}_{-n/w}$. Indicating its charges by
\begin{equation}
    (\gen{L}_0,-\gen{J}^3;\ \genr{L}_0,-\genr{J}^3)\,,
\end{equation}
we have seen that we have, at $h>0$
\begin{equation}
    T^{A\dot{A}}(p)|0\rangle_w\sim \alpha^{A\dot{A}}_{-\frac{n}{w}}|0\rangle_w
    \sim \Big(
\frac{\delta H+\tfrac{1}{2\pi}p+w}{2},-\frac{w}{2};\ 
\frac{\delta H-\tfrac{1}{2\pi}p+w}{2},-\frac{w}{2}
\Big),
\end{equation}
where $\delta H$ is the anomalous dimension due to the RR deformation. In particular
\begin{equation}
    \delta H=\sqrt{\frac{1}{4\pi^2}p^2+4h^2\sin^2(\frac{p}{2})}-\frac{1}{2\pi}p=O(h^2)\,.
\end{equation}
Suppose that we now act with a chiral charge, such as $\gen{Q}^{B}\varepsilon_{AB}$. Then, it is clear that we get the fermion excitation with the same $H$ and $M$ (equivalently, same $L_0-J^3$ and $\tilde{L}_0-\tilde{J}^3$), but with shifted $L_0$ and $J^3$ charges (both increased):
\begin{equation}
    \Psi^{+\dot{A}}(p)|0\rangle_w\sim \Psi^{+\dot{A}}_{-\frac{1}{2}-\frac{n}{w}}|0\rangle_w
    \sim \Big(
\frac{\delta H+\tfrac{1}{2\pi}p+w+1}{2},-\frac{w+1}{2};\ 
\frac{\delta H-\tfrac{1}{2\pi}p+w}{2},-\frac{w}{2}
\Big).
\end{equation}
Suppose now that we act with the antichiral charge $\genr{S}^{+B}\varepsilon_{AB}$. According to figure~\ref{fig:representations} we expect this action to give the same state~$\Psi^{+\dot{A}}(p)$ in the decompactification ($w\to\infty$) limit. However, the superconformal charge $\genr{S}^{+B}\varepsilon_{AB}$ should decrease both $\tilde{L}_0$ and $\tilde{J}^3$, and give 
\begin{equation}
    \Big(
\frac{\delta H+\tfrac{1}{2\pi}p+w}{2},-\frac{w}{2};\ 
\frac{\delta H-\tfrac{1}{2\pi}p+w-1}{2},-\frac{w-1}{2}
\Big).
\end{equation}
The correct way to interpret this is in terms of a \textit{length-changing effect}. This intuition is behind the famous dynamical spin-chain of Beisert~\cite{Beisert:2005tm} and it is also the way that integrability was first understood in the context of $AdS_3/CFT_2$~\cite{Borsato:2012ud}. Namely we have the state
\begin{equation}
\begin{aligned}
&\Psi^{+\dot{A}}(p)|0\rangle_{w-1}\sim \Psi^{+\dot{A}}_{-\frac{1}{2}-\frac{n}{w-1}}|0\rangle_{w-1}\\
    &\qquad\sim
    \Big(
\frac{\delta H+\tfrac{1}{2\pi}p+(w-1)+1}{2},-\frac{(w-1)+1}{2};\ 
\frac{\delta H-\tfrac{1}{2\pi}p+w-1}{2},-\frac{w-1}{2}
\Big).
\end{aligned}
\end{equation}
If we had considered instead the action of $\genr{Q}^{A}$, which is related to $\gen{S}^A$, we would have found instead $w\to w+1$.
In a similar way, we find that the action of the chiral generators on anti-chiral states must also change the length $w\to w\pm1$.
All these expressions are apparently problematic, because our modes are originally quantised in $1/w$ rather than in $1/(w\pm1)$; however, this is just a sign that this action is strictly speaking only well defined in the $w\to\infty$ limit, as  is well known~\cite{Arutyunov:2006yd}.

Of course we know that this action is zero at order~$h^0$: at that order, only the chiral-on-chiral and antichiral-on-antichiral actions are non-vanishing. In fact, an explicit expansion of the representation coefficients~\eqref{eq:reprcoefficients}
\begin{equation}
\begin{aligned}
    a(p)=a_0(p)+h\,\delta a(p)+O(h^2)\,,\quad
    \bar{a}(p)=\bar{a}_0(p)+h\,\delta \bar{a}(p)+O(h^2)\,,\\
    b(p)=b_0(p)+h\,\delta b(p)+O(h^2)\,,\quad
    \bar{b}(p)=\bar{b}_0(p)+h\,\delta \bar{b}(p)+O(h^2)\,,
\end{aligned}
\end{equation}
shows that the chiral-on-antichiral and antichiral-on-chiral actions are of order~$\mathcal{O}(h)$. More specifically we have at order~$h$ for the antichiral-on-chiral
\begin{equation}
\label{eq:antichiral-on-chiral}
\begin{aligned}
    \genr{Q}^A \alpha_{-\frac{n}{w}}^{B\dot{A}}|0\rangle_{w}&=
    h\,\varepsilon^{AB}\delta b(p)\, \Psi^{-\dot{A}}_{+\frac{1}{2}-\frac{n}{w+1}}|0\rangle_{w+1},\\
    \genr{Q}^A \Psi^{+\dot{A}}_{-\frac{1}{2}-\frac{n}{w}}|0\rangle_{w}&=
    h\,\delta b(p)\, \alpha^{A\dot{A}}_{-\frac{n}{w+1}}|0\rangle_{w+1},
    \\
    \genr{S}^A \alpha_{-\frac{n}{w}}^{B\dot{A}}|0\rangle_{w}&=
    h\,\varepsilon^{AB}\delta \bar{b}(p)\, \Psi^{+\dot{A}}_{-\frac{1}{2}-\frac{n}{w-1}}|0\rangle_{w-1},\\
    \genr{S}^A \Psi^{-\dot{A}}_{+\frac{1}{2}-\frac{n}{w}}|0\rangle_{w}&=
    h\,\delta \bar{b}(p)\, \alpha^{A\dot{A}}_{-\frac{n}{w-1}}|0\rangle_{w-1},
\end{aligned}
\end{equation}
and for the chiral-on-antichiral
\begin{equation}
\label{eq:chiral-on-antichiral}
\begin{aligned}
    \gen{Q}^A \tilde{\alpha}_{-\frac{n}{w}}^{B\dot{A}}|0\rangle_{w}&=
    h\,\varepsilon^{AB}\delta a(p)\, \tilde\Psi^-_{+\frac{1}{2}-\frac{n}{w+1}}|0\rangle_{w+1},\\
    \gen{Q}^A \tilde\Psi^+_{-\frac{1}{2}-\frac{n}{w}}|0\rangle_{w}&=
    h\,\delta a(p)\,\tilde{\alpha}_{-\frac{n}{w+1}}^{A\dot{A}} |0\rangle_{w+1},\\
    \gen{S}^A \tilde{\alpha}_{-\frac{n}{w}}^{B\dot{A}}|0\rangle_{w}&=
    h\,\varepsilon^{AB}\delta \bar{a}(p)\, \tilde\Psi^+_{-\frac{1}{2}-\frac{n}{w-1}}|0\rangle_{w-1},\\
    \gen{S}^A \tilde\Psi^+_{-\frac{1}{2}-\frac{n}{w}}|0\rangle_{w}&=
    h\,\delta \bar{a}(p)\,\tilde{\alpha}_{-\frac{n}{w-1}}^{A\dot{A}} |0\rangle_{w-1}.
\end{aligned}
\end{equation}
The explicit form of the first-order representation coefficients can also be found from Taylor-expanding those of~\cite{Lloyd:2014bsa}. We find
\begin{equation}
\begin{aligned}
        a(p)=&
    \begin{cases}
    +e^{+\frac{i}{4}p+i\xi(p)}\sqrt{\frac{p}{2\pi}}+O(h^2)&\text{if}\quad p>0\,,\\[0.4cm]
    \displaystyle s\,e^{+\frac{i}{4}p+i\xi(p)}\frac{\sin(\frac{p}{2})}{\sqrt{-\frac{p}{2\pi}}}\,h+O(h^3)&\text{if}\quad p<0\,,
    \end{cases}\\[0.2cm]
        \bar{a}(p)=&
    \begin{cases}
    +e^{-\frac{i}{4}p-i\xi(p)}\sqrt{\frac{p}{2\pi}}+O(h^2)&\text{if}\quad p>0\,,\\[0.4cm]
    \displaystyle s\,e^{-\frac{i}{4}p-i\xi(p)}\frac{\sin(\frac{p}{2})}{\sqrt{-\frac{p}{2\pi}}}\,h+O(h^3)&\text{if}\quad p<0\,,
    \end{cases}\\[0.2cm]
\end{aligned}
\end{equation}
while for the remaining coefficients the role of negative and positive momenta swaps,
\begin{equation}
\begin{aligned}
    b(p)&=
    \begin{cases}
    \displaystyle -e^{+\frac{i}{4}p+i\xi(p)}\frac{\sin(\frac{p}{2})}{\sqrt{\frac{p}{2\pi}}}\,h+O(h^3)&\text{if}\quad p>0\,,\\[0.7cm]
    s\,e^{+\frac{i}{4}p+i\xi(p)}\sqrt{-\frac{p}{2\pi}}+O(h^2)&\text{if}\quad p<0\,,
    \end{cases}\\
    \bar{b}(p)&=
    \begin{cases}
    \displaystyle -e^{-\frac{i}{4}p-i\xi(p)}\frac{\sin(\frac{p}{2})}{\sqrt{\frac{p}{2\pi}}}\,h+O(h^3)&\text{if}\quad p>0\,,\\[0.7cm]
    s\,e^{-\frac{i}{4}p-i\xi(p)}\sqrt{-\frac{p}{2\pi}}+O(h^2)&\text{if}\quad p<0\,,
    \end{cases}
\end{aligned}
\end{equation}
In the language of~\cite{Sfondrini:2014via} (see in particular section~5 there), the positive-momentum (chiral) representations behave like ``left'' representations, and the negative-momentum (antichiral) ones behave like ``right'' representations.

\subsection{The ``spin-chain frame'' from the length-changing effects}
If we take the view that the length-changing effects only occur for antichiral-on-chiral or chiral-on-antichiral actions, it is natural to ask the coproduct to be trivial for the chiral-on-chiral or antichiral-on-antichiral action. This is the \textit{spin-chain frame} coproduct, which we define separately for the two representations.
In this context, the braiding factors in the coproduct have the natural interpretation of $\mathcal{Z}^\pm$-markers~\cite{Beisert:2005tm}, see also section~5 of~\cite{Sfondrini:2014via}, which have a Zamolodchikov-Faddev algebra of sorts,
\begin{equation}
\label{eq:braiding}
    \mathcal{Z}^\pm \mathcal{X}(p)=e^{\mp ip}\mathcal{X}(p)\mathcal{Z}^\pm\,,
\end{equation}
where $\mathcal{X}(p)$ is any excitation.
Hence, in this frame it is also natural to set
\begin{equation}
    \xi(p)=0\,.
\end{equation}

\paragraph{Chiral representation ($p>0$).}
In this case we want the chiral supercharge to have a trivial coproduct, and the antichiral ones to have a nontrivial braiding. Hence we set:
\begin{equation}
\label{eq:scframe1}
\begin{aligned}
    Q^A(p_1,p_2)&=Q^A(p_1)\otimes 1\quad&&+\Sigma\otimes Q^{A}(p_2)\,,\\
    \widetilde{Q}^A(p_1,p_2)&=\widetilde{Q}^A(p_1)\otimes 1\,e^{+ip_2}&&+\Sigma\otimes \widetilde{Q}^{A}(p_2)\,,\\
    S^A(p_1,p_2)&=S^A(p_1)\otimes 1\quad&&+\Sigma\otimes S^{A}(p_2)\,,\\
    \widetilde{S}^A(p_1,p_2)&=\widetilde{S}^A(p_1)\otimes 1\,e^{-ip_2}&&+\Sigma\otimes \widetilde{S}^{A}(p_2)\,.
\end{aligned}
\end{equation}

\paragraph{Antichiral representation ($p<0$).}
In a similar (or rather opposite) way,  we set:
\begin{equation}
\label{eq:scframe2}
\begin{aligned}
    Q^A(p_1,p_2)&=Q^A(p_1)\otimes 1\,e^{+ip_2}&&+\Sigma\otimes Q^{A}(p_2)\,,\\
    \widetilde{Q}^A(p_1,p_2)&=\widetilde{Q}^A(p_1)\otimes 1&&+\Sigma\otimes \widetilde{Q}^{A}(p_2)\,,\\
    S^A(p_1,p_2)&=S^A(p_1)\otimes 1\,e^{-ip_2}&&+\Sigma\otimes S^{A}(p_2)\,,\\
    \widetilde{S}^A(p_1,p_2)&=\widetilde{S}^A(p_1)\otimes 1&&+\Sigma\otimes \widetilde{S}^{A}(p_2)\,,
\end{aligned}
\end{equation}

\paragraph{String-frame coproduct.}
For the reader's convenience we repeat here the string frame coproduct~\eqref{eq:coproduct-string} in the simpler case $\xi(p)=0$. Note how it is sort of a ``symmetrised'' version of the spin-chain frame one.%
\footnote{
The fact that the braiding factor is on the second term is a matter of convention, though not inconsequential; technically, the coproduct in the spin-chain frame is the opposite of the one in the string frame.
}
\begin{equation}
\label{eq:stringframe}
\begin{aligned}
    Q^A(p_1,p_2)&=Q^A(p_1)\otimes 1+e^{+\tfrac{i}{2}p_1}\,\Sigma\otimes Q^{A}(p_2)\,,\\
    \widetilde{Q}^A(p_1,p_2)&=\widetilde{Q}^A(p_1)\otimes 1+e^{+\tfrac{i}{2}p_1}\,\Sigma\otimes \widetilde{Q}^{A}(p_2)\,,\\
    S^A(p_1,p_2)&=S^A(p_1)\otimes 1+e^{-\tfrac{i}{2}p_1}\,\Sigma\otimes S^{A}(p_2)\,,\\
    \widetilde{S}^A(p_1,p_2)&=\widetilde{S}^A(p_1)\otimes 1+e^{-\tfrac{i}{2}p_1}\,\Sigma\otimes \widetilde{S}^{A}(p_2)\,,
\end{aligned}
\end{equation}

\section{Comparison with the results of Gaberdiel-Gopakumar-Nairz}
In~\cite{Gaberdiel:2023lco} the authors have compute the length-changing action of eqs.~\eqref{eq:antichiral-on-chiral} and \eqref{eq:chiral-on-antichiral} starting within the framework of conformal perturbation theory. Technically, this is the computation of a correlation function involving the states on the r.h.s.~of eqs.~\eqref{eq:antichiral-on-chiral} and \eqref{eq:chiral-on-antichiral} (say at the origin of the complex plane), those on the l.h.s.\ (say at infinity), in the presence of the supercharges and of the deformation operator~$\mathcal{O}$ of eq.~\eqref{eq:deformation} (at first-order in~$h$).
In fact, they also perform the computation in the presence of \textit{two} magnon-like excitations, which allows them to extract the non-trivial coproduct. Their computation is performed at finite~$w$, but as discussed we are only interested in the $w\to\infty$ limit of their results, because only in this decompactification limit we can truly give a sense to this algebra.

Looking at the form of the results of~\cite{Gaberdiel:2023lco}, we first observe a slightly different notation with the one used in the integrability literature (and in the sections above). Namely the definition of the momentum is related as  follows
\begin{equation}
\label{eq:GGNmomentum}
    \big(1-p\big)\Big|_{\text{GGN}} = \frac{p}{2\pi}\Big|_{\text{here}}.
\end{equation}
Translating their results in our notation, we notice that   the charges $\gen{Q}$ which for us and in the $AdS_3$ integrability literature are the supersymmetry generators (\textit{i.e.}, their anticommutator gives a derivative) are called~$S$ in~\cite{Gaberdiel:2023lco}. The simplest way to account for such a discrepancy is to relabel the highest- and lowest-weight states of each representation (which is tantamount to exchanging lowering and raising operators).
With this in mind we see that on the one-particle representation they find
\begin{equation}
    C(p)=\frac{g}{2i}\big(1-e^{-ip}\big)\,.
\end{equation}
This indeed coincides with our~\eqref{eq:centralchargesp} for
\begin{equation}
   \xi(p)=0\,, 
\end{equation}
 with the identification
\begin{equation}
    h=g\,.
\end{equation}
Hence, this allows to match the parameter~$h$, which in~\cite{Lloyd:2014bsa} was introduced just on the basis of the allowed symmetries, with the symmetric-product orbifold parameter~$g$. Of course we expected the relation to be linear, but it is interesting to note that the normalisation coefficient seems to be precisely one.

We can now look at the representation coefficient; already from section 3.3 of~\cite{Gaberdiel:2023lco}, which deals with~$O(h^0)$, we notice a discrepancy: there, the coefficients are chosen to be real (as a consequence of the definition of the supercharges), whereas what we get from expanding~\cite{Lloyd:2014bsa}, \eqref{eq:reprcoefficientshiszero} is not real for real momentum. This is of course not an issue, as the precise form of the representation depends on the choice of a basis. It is sufficient to rescale our fermions to get rid of the complex terms in~\eqref{eq:reprcoefficientshiszero}. To this end, it is sufficient to rescale
\begin{equation}
\label{eq:change1part}
    \chi(p)\to e^{+\frac{i}{4}|p|}\chi(p)\,,\qquad
    \alpha^{A\dot{A}}(p)\to\alpha^{A\dot{A}}(p),\qquad
    \tilde{\chi}(p)\to e^{-\frac{i}{4}|p|}\tilde{\chi}(p)\,.
\end{equation}
In fact, we can also get rid of the fastidious sign~$s=\pm1$ of~\eqref{eq:reprcoefficientshiszero} by this rescaling.
This rescaling can be rewritten in terms of the symmetric-orbifold excitations using~\eqref{eq:chiralident} and~\eqref{eq:antichiralident}. In that language, the appearance  of~$|p|$ in the exponent is not concerning at all, as we anyway treat $p>0$ and $p<0$ as separate representations.

After the rescaling of the fermions to make the $O(h^0)$ term real, we see that the $O(h^1)$ picks up a phase of the type $e^{\pm\frac{i}{2}p}\sin(p/2)=\pm(1-e^{\pm ip})/(2i)$, so that we get%
\footnote{Note that we would find the same expansion at this order even if the dependence on $h$ were given by a functions~$f(h)=h+O(h^3)$.}
\begin{equation}
\begin{aligned}
        a(p)=&
    \begin{cases}
    +\sqrt{\frac{p}{2\pi}}+O(h^2)&\text{if}\quad p>0\,,\\[0.4cm]
    \displaystyle +\frac{1-e^{-ip}}{\sqrt{-\frac{p}{2\pi}}}\,\frac{h}{2i}+O(h^3)&\text{if}\quad p<0\,,
    \end{cases}\\[0.2cm]
        \bar{a}(p)=&
    \begin{cases}
    +\sqrt{\frac{p}{2\pi}}+O(h^2)&\text{if}\quad p>0\,,\\[0.4cm]
    \displaystyle -\frac{1-e^{+ip}}{\sqrt{-\frac{p}{2\pi}}}\,\frac{h}{2i}+O(h^3)&\text{if}\quad p<0\,,
    \end{cases}\\[0.2cm]
\end{aligned}
\end{equation}
and similarly
\begin{equation}
\begin{aligned}
    b(p)&=
    \begin{cases}
    \displaystyle +\frac{1-e^{-ip}}{\sqrt{\frac{p}{2\pi}}}\,\frac{h}{2i}+O(h^3)&\text{if}\quad p>0\,,\\[0.7cm]
    +\sqrt{-\frac{p}{2\pi}}+O(h^2)&\text{if}\quad p<0\,,
    \end{cases}\\
    \bar{b}(p)&=
    \begin{cases}
    \displaystyle -\frac{1-e^{+ip}}{\sqrt{\frac{p}{2\pi}}}\,\frac{h}{2i}+O(h^3)&\text{if}\quad p>0\,,\\[0.7cm]
    +\sqrt{-\frac{p}{2\pi}}+O(h^2)&\text{if}\quad p<0\,.
    \end{cases}
\end{aligned}
\end{equation}
These appear to be the representation coeffiecients of~\cite{Gaberdiel:2023lco}. It is of course unsurprising that we can match the representation coefficients up to a change of basis: since we are dealing with two irreducible representations which have the same central charges, they must be unitarily equivalent.
By picking a basis in the representation space, we can represent the change of basis~\eqref{eq:change1part} as a diagonal matrix $\gen{V}(p)$, which acts on any state~$\vec{\Phi}(p)$ as
\begin{equation}
    \vec{\Phi}(p)\to\gen{V}(p)\cdot\vec{\Phi}(p)\,.
\end{equation}

It remains to deal with the co-product of~\cite{Gaberdiel:2023lco}, which is in the ``spin-chain frame'' rather than in the ``string frame''~\cite{Lloyd:2014bsa}, see above for a discussion of such frames, and in particular eqs.~\eqref{eq:scframe1} to~\eqref{eq:stringframe}. The map between these two frames is described in detail in~\cite{Sfondrini:2014via}, starting from eqs.~(5.65). Let us briefly summarise it. We start by making the spin-chain frame coproduct a little more symmetric, so that it has the same form for either representation. We perform a change of basis of the \textit{two-particle} Hilbert space by means of a matrix
\begin{equation}
    \gen{U}(p_2)\otimes 1\,,
\end{equation}
Note that this change of basis cannot be induced by a change of the one-particle basis. A diagonal matrix is sufficient to obtain
\begin{equation}
\begin{aligned}
    U^\dagger_{p_2}\otimes 1\cdot Q^A(p_1,p_2)\cdot U_{p_2}\otimes 1&=Q^A(p_1)\otimes 1\,e^{+\frac{i}{2}p_2}+\Sigma\otimes Q^{A}(p_2)\,,\\
    U^\dagger_{p_2}\otimes 1\cdot \widetilde{Q}^A(p_1,p_2)\cdot U_{p_2}\otimes 1&=\widetilde{Q}^A(p_1)\otimes 1\,e^{+\frac{i}{2}p_2}+\Sigma\otimes \widetilde{Q}^{A}(p_2)\,,\\
    U^\dagger_{p_2}\otimes 1\cdot S^A(p_1,p_2)\cdot U_{p_2}\otimes 1&=S^A(p_1)\otimes 1\,e^{-\frac{i}{2}p_2}+\Sigma\otimes S^{A}(p_2)\,,\\
    U^\dagger_{p_2}\otimes 1\cdot \widetilde{S}^A(p_1,p_2)\cdot U_{p_2}\otimes 1&=\widetilde{S}^A(p_1)\otimes 1\,\,e^{-\frac{i}{2}p_2}+\Sigma\otimes \widetilde{S}^{A}(p_2)\,,
\end{aligned}
\end{equation}
which is reminiscent of the string-frame coproduct up to having the braiding on the first, rather than second, term. In fact, this is the \textit{opposite coproduct} with respect to what we want. If we introduce the labels ``s.c.'' and ``str.'' we have that schematically
\begin{equation}
\label{eq:Q12relations}
    U^\dagger_{p_2}\otimes 1\cdot Q^A_{\text{s.c.}}(p_1,p_2)\cdot U_{p_2}\otimes 1 = \Pi^g\cdot Q^A_{\text{str.}}(p_2,p_1)\cdot\Pi^g\,,
\end{equation}
where $\Pi^g$ is the graded permutation matrix which swaps the first and the second Hilbert spaces (taking also care of the Fermion signs). 

Recall~\cite{Lloyd:2014bsa} that the S-matrix~$\mathcal{S}(p_1,p_2)$ is uniquely fixed, up to a dressing factor for each choice of irreducible representations, by the symmetries which we have been discussing. Let us say that in string-frame we have
\begin{equation}
    Q(p_2,p_1)\cdot \mathcal{S}(p_1,p_2)=\mathcal{S}(p_1,p_2)\cdot Q(p_1,p_2)\,.
\end{equation}
Hence, due to~\eqref{eq:Q12relations}, we have that the two S~matrices are also related:
\begin{equation}
\label{eq:changeofbasis}
    \mathcal{S}_{\text{str.}}(p_1,p_2)=
    \Pi^g\cdot 1\otimes U^\dagger_{p_1}\cdot V_{p_2}^\dagger\otimes V_{p_1}^\dagger\cdot
    \mathcal{S}_{\text{s.c.}}(p_2,p_1)\cdot V_{p_2}\otimes V_{p_1}\cdot U_{p_1}\otimes 1\cdot\Pi^g\,,
\end{equation}
where the matrices~$V$ and~$U$ are actually diagonal.

\paragraph{Notation for symmetry generators and states.}

In this paragraph we relate the notation for the symmetry generators used in the integrability literature (which  we have used above) with that of~\cite{Gaberdiel:2023lco}. Let us start with the identification of the symmetry generators. Above, see in particular eqs.~\eqref{eq:superchargesnotation} and~\eqref{eq:chargesthatannihilate}, we have denoted by~$\gen{Q}$ the supersymmetry generators (which have positive dimension under the dilatation operator, and whose anticommutator yields  translations) and with~$\gen{S}$ the superconformal symmetry generator (which have negative dimension under the dilatation operator, and whose anticommutator yields the special conformal transformations); this follows the integrability literature (see in particular~\cite{Lloyd:2014bsa} and, for a more detailed discussion of the algebra,~\cite{Eden:2021xhe}).
The notation of~\cite{Gaberdiel:2023lco} is different and we have
\begin{equation}
\begin{aligned}
Q_1\Big|_{\text{there}}\!\!\!= \gen{S}_{2}\Big|_{\text{here}},\quad
Q_2\Big|_{\text{there}}\!\!\!= -\gen{S}_{1}\Big|_{\text{here}},\quad
S_1\Big|_{\text{there}}\!\!\!= \gen{Q}^{2}\Big|_{\text{here}},\quad
S_2\Big|_{\text{there}}\!\!\!= -\gen{Q}^{1}\Big|_{\text{here}},\\
\widetilde{Q}_1\Big|_{\text{there}}\!\!\!= -\genr{S}^{1}\Big|_{\text{here}},\quad
\widetilde{Q}_2\Big|_{\text{there}}\!\!\!= \genr{S}^{2}\Big|_{\text{here}},\quad
\widetilde{S}_1\Big|_{\text{there}}\!\!\!= -\genr{Q}_{1}\Big|_{\text{here}},\quad
\widetilde{S}_2\Big|_{\text{there}}\!\!\!= \genr{Q}_{2}\Big|_{\text{here}}.
\end{aligned}
\end{equation}
In  the integrability notation, the indices on the supercharges indicate their transformation under the $\su(2)_{\bullet}$ automorphism and are raised and lowered with the Levi-Civita tensor.
As a result, the identification of the states for the chiral representation (that is, for the representation that at $h=0$ is annihilated by the $\genr{Q}$ and $\genr{S}$ charges) is
\begin{equation}
    \psi^{-}\Big|_{\text{there}}\!\!\!= 
    \chi^{\dot{1}}\Big|_{\text{here}},\qquad
    \alpha^{A}\Big|_{\text{there}}\!\!\!= 
    T^{A\dot{1}}\Big|_{\text{here}},\quad(A=1,2),\qquad
    \psi^{+}\Big|_{\text{there}}\!\!\!= 
    \tilde{\chi}^{\dot{1}}\Big|_{\text{here}}\,,
\end{equation}
where the momentum of the magnons for the chiral representation is identified with the mode number as explained above, see in particular Table~\ref{tab:cftcartan}.
The index $\dot{A}=\dot{1}$ refers to $\su(2)_\circ$. There is another identical representation,  with $\dot{A}=\dot{2}$, which in terms of~\cite{Gaberdiel:2023lco} is related to the excitations~$\bar{\alpha}^{A}$ ($A=1,2$) and~$\bar{\psi}^\pm$.
As for the anti-chiral representation we have
\begin{equation}
    \tilde\psi^{+}\Big|_{\text{there}}\!\!\!= 
    \chi^{\dot{1}}\Big|_{\text{here}},\qquad
    \tilde\alpha^{A}\Big|_{\text{there}}\!\!\!= 
    T^{A\dot{1}}\Big|_{\text{here}},\quad(A=1,2),\qquad
    \tilde\psi^{-}\Big|_{\text{there}}\!\!\!= 
    \tilde{\chi}^{\dot{1}}\Big|_{\text{here}}\,,
\end{equation}
where now \emph{the relation with mode number and momenta has the opposite sign than in the chiral representation}, cf.~again Table~\ref{tab:cftcartan}. 
Note that the highest-weight state in the representation, corresponding to $\chi^{\dot{1}}$, is $\tilde\psi^{+}$ rather than $\tilde\psi^{-}$, because spin $\gen{J}^3-\genr{J}^3$ has to match (while the total R-charge $\gen{R}$ is decoupled at $w=\infty$).
Finally, the anti-chiral representation with $\dot{A}=\dot{2}$ is related to the excitations~$\tilde{\bar{\alpha}}^{A}$ ($A=1,2$) and~$\tilde{\bar{\psi}}^\pm$ of~\cite{Gaberdiel:2023lco}. 
With this identification of the states, and keeping in mind the convention of~\cite{Gaberdiel:2023lco} for the momentum~\eqref{eq:GGNmomentum}, the matching of the S~matrix presented in~\cite{Gaberdiel:2023lco} with the one existing in the integrability literature, e.g.~\cite{Lloyd:2014bsa,Eden:2021xhe}
follows from the change of basis~\eqref{eq:changeofbasis}.

\subsection{Further details on the dictionary between spin-chain and string frames}

After this paper appeared on the arXiv, the authors of~\cite{Gaberdiel:2023lco} revised their manuscript but maintained that the S~matrix which they constructed may be different in the LR/RL sector. We find it therefore appropriate to further clarify this point.
One possible cause for confusion%
\footnote{We thank the anonymous referee for bringing this point to our attention.}
may be that in the LR S-matrix in the spin-chain frame it is necessary to introduce Beisert's $\mathcal{Z}^\pm$ markers, while this is not so in the string-frame S~matrix.
While this has been discussed elsewhere at some length~\cite{Arutyunov:2006yd,Arutyunov:2009ga,Sfondrini:2014via}, let us further comment on the case at hand.

To begin with, it may be worth noting that the early literature for $AdS_3\times S^3\times T^4$ integrability has actually been written in the spin-chain frame, using the $\mathcal{Z}^\pm$ markers, see for instance eqs.~(3.10) and~(3.13) in~\cite{Borsato:2013qpa}; in the same paper, the equivalence with the string-frame is discussed. Ref.~\cite{Borsato:2013qpa} dealt with the pure-RR massive S-matrix, but actually the discussion of the spin-versus-string frame is the same that of this case, because it only hinges on length-changing effects (the shifts of $\gen{R}$-charge) as we discussed in Section~\ref{sec:orbifold:boundstates}, and these are the same as here.
The reason while most of the subsequent integrability literature has been directly written in the string-frame is that this allowed to straightforwardly match with many near-pp-wave computations developed over the years, see e.g.~\cite{Sundin:2016gqe} and references therein.

Let us now elaborate a bit on the spin-chain frame. If we imagine to have a spin chain of large but finite length~$w$, we find that it is necessary to shift $w\to w\pm1$ to correctly account for the action of, say, a chiral generator on an anti-chiral excitation, as we have discussed in Section~\ref{sec:orbifold:boundstates}. What we have represented there in terms of shifts of~$w\to w\pm1$ may also be represented by inserting the~$\mathcal{Z}^\pm$ markers which insert or remove one site close to the magnon (let's say, by convention, immediately to its right).
In this language, for instance, equation~\eqref{eq:antichiral-on-chiral} for the chiral representation would read
\begin{equation}
\label{eq:zmarkersRL1}
\begin{aligned}
    \genr{Q}^A |T_p^{B\dot{A}}\rangle&=
    \varepsilon^{AB} b(p)\; |\chi_p^{\dot{A}}\mathcal{Z}^+\rangle,\qquad
    &\genr{Q}^A |\tilde{\chi}^{\dot{A}}_{p}\rangle&=
     b(p)\; |T^{A\dot{A}}_{p}\mathcal{Z}^+\rangle,
    \\
    \genr{S}^A |T_{p}^{B\dot{A}}\rangle&=
    \varepsilon^{AB} \bar{b}(p)\; |\tilde{\chi}^{\dot{A}}_{p}\mathcal{Z}^-\rangle,\qquad
    &\genr{S}^A |\chi^{\dot{A}}_{p}\rangle&=
     \bar{b}(p)\; |T^{A\dot{A}}_{p}\mathcal{Z}^-\rangle,
\end{aligned}
\end{equation}
where to further simplify the formulae we have already made the identifications of Table~\ref{tab:cftcartan} and traded the mode number $n$ for the momentum $p=2\pi n/w$; we have also reinstated $b(p)=h\,\delta b(p)+O(h^2)$ and  $\bar{b}(p)=h\,\delta \bar{b}(p)+O(h^2)$ because the small-$h$ expansion is actually irrelevant (and distracting) for the point we are making.
On top of these relations, we also have the action of the chiral generators on the chiral representation, which does not include any length-changing effects,
\begin{equation}
\label{eq:zmarkersRL2}
\begin{aligned}
    \gen{Q}^A |T_{p}^{B\dot{A}}\rangle&=
    \varepsilon^{AB} a(p)\; |\tilde{\chi}^{\dot{A}}_{p}\rangle,\qquad
    &\gen{Q}^A |\chi^{\dot{A}}_{p}\rangle&=
     \a(p)\; |T^{A\dot{A}}_{p}\rangle,\\
    \gen{S}^A |T_p^{B\dot{A}}\rangle&=
    \varepsilon^{AB} \bar{a}(p)\; |\chi_p^{\dot{A}}\rangle,\qquad
    &\gen{S}^A |\tilde{\chi}^{\dot{A}}_{p}\rangle&=
     \bar{a}(p)\; |T^{A\dot{A}}_{p}\rangle.
\end{aligned}
\end{equation}
The only way to interpret eqs.~\eqref{eq:zmarkersRL1} and~\eqref{eq:zmarkersRL2} as a representation is to identify
\begin{equation}
\label{eq:zmarkerid1}
    |\chi_p^{\dot{A}}\mathcal{Z}^\pm\rangle=|\chi_p^{\dot{A}}\rangle,\quad
    |\chi_p^{\dot{A}}\mathcal{Z}^\pm\rangle=|\chi_p^{\dot{A}}\rangle,\quad
    |\tilde{\chi}_p^{\dot{A}}\mathcal{Z}^\pm\rangle=|\tilde{\chi}_p^{\dot{A}}\rangle,\qquad \text{as}\quad w\to\infty\,.
\end{equation}
This is absolutely fine in the decompactification limit, as  $\gen{R}$ decouples. Then one has a four dimensional module (for each value of $\dot{A}=1,2$) of the type of Figure~\ref{fig:representations}, as it should be.

This makes it seem that the length-changing effects play no role in the above representation, and that is actually true \textit{for the one-magnon representation}. It is however not true for the multi-magnon representation. For instance, on a two-magnon state one would get expressions of the form
\begin{equation}
    \genr{Q}^A |\tilde{\chi}^{\dot{A}_1}_{p_1}\tilde{\chi}^{\dot{A}_2}_{p_2}\rangle=
     b(p_1)\; |T^{A\dot{A}_1}_{p_1}\mathcal{Z}^+\tilde{\chi}^{\dot{A}_2}_{p_2}\rangle-
     b(p_2)\; |\tilde{\chi}^{\dot{A}_1}_{p_1}T^{A\dot{A}_2}_{p_2}\mathcal{Z}^+\rangle\,,
\end{equation}
and so on. For the chiral-chiral two-particle representation, instead of a four-dimensional module, we get sixteen states, with various insertions of $Z^\pm$ either after the first or the second excitation. Again, in the $w\to\infty$ limit this is just a sixteen-dimensional module, thanks to identifications of the type
\begin{equation}
\label{eq:zmarkerid2}
    |T^{A\dot{A}_1}_{p_1}\mathcal{Z}^\pm\tilde{\chi}^{\dot{A}_2}_{p_2}\rangle=
    e^{\mp i p_2}\,
    |T^{A\dot{A}_1}_{p_1}\tilde{\chi}^{\dot{A}_2}_{p_2}\mathcal{Z}^+\rangle
    = e^{\mp i p_2}\,|T^{A\dot{A}_1}_{p_1}\tilde{\chi}^{\dot{A}_2}_{p_2}\rangle\,.
\end{equation}
More generally, if $\mathcal{X}_p$ is any magnon, one accounts for the effect of adding or subtracting a site like we mentioned in~\eqref{eq:braiding}, that is
\begin{equation}
\label{eq:zmarkerid3}
    |\mathcal{Z}^\pm\mathcal{X}_p\rangle=
    e^{\mp i p}\,|\mathcal{X}_p\mathcal{Z}^\pm\rangle= e^{\mp i p}\,|\mathcal{X}_p\rangle\,,\qquad\text{as}\quad w\to\infty\,.
\end{equation}
The net effect of the bookkeeping provided by the~$\mathcal{Z}^\pm$ markers is then to endow the two-particle representation with a non-trivial coproduct in terms of the braiding factors~$e^{\pm i p_2}$, \textit{e.g.},
\begin{equation}
\begin{aligned}
    \genr{Q}^A |\tilde{\chi}^{\dot{A}_1}_{p_1}\tilde{\chi}^{\dot{A}_2}_{p_2}\rangle&=
     e^{ip_2}b(p_1)\; |T^{A\dot{A}_1}_{p_1}\tilde{\chi}^{\dot{A}_2}_{p_2}\mathcal{Z}^+\rangle-
     b(p_2)\; |\tilde{\chi}^{\dot{A}_1}_{p_1}T^{A\dot{A}_2}_{p_2}\mathcal{Z}^+\rangle\,,\\
     &=
     e^{ip_2}b(p_1)\; |T^{A\dot{A}_1}_{p_1}\tilde{\chi}^{\dot{A}_2}_{p_2}\rangle-
     b(p_2)\; |\tilde{\chi}^{\dot{A}_1}_{p_1}T^{A\dot{A}_2}_{p_2}\rangle\,,
\end{aligned}
\end{equation}
where in the last line we omitted the $\mathcal{Z}^+$-marker as it is irrelevant in the decompactification limit. This gives precisely the representaiton of eq.~\eqref{eq:scframe1}.
The story is the same for the antichiral-antichiral representation, see eq.~\eqref{eq:scframe2}, as well as for the chiral-antichiral and antichiral-chiral, which follow straightforwardly --- because the braiding factors $e^{\pm i p_2}$ depend only on the $\mathcal{Z}^\pm$ marker arising from the first particle, the coproduct of the chiral-antichiral representation takes the form~\eqref{eq:scframe1} and the coproduct of the antichiral-chiral representation takes the form~\eqref{eq:scframe2}. Of course one should use the appropriate form of the one-particle representation  coefficients $Q^A(p_j)$, $\tilde{Q}^A(p_j)$, $S^A(p_j)$, $\tilde{S}^A(p_j)$ depending on whether $p_j$ is chiral or antichiral.
In any case, the form of these coproducts and the map between string frame and spin-chain frame was worked out and discussed at some lenght in the literature, see e.g.\  eq.~(5.44) of~\cite{Sfondrini:2014via}. Let us provide some guidance on how to read those results. Briefly: it is actually convenient to first use the fact that the one-particle irreducible representations that we are considering may be written as tensor products of two-dimensional representations of a smaller algebra generated by four supercharges $\gen{Q},$ $\genr{Q},$ $\gen{S},$ $\genr{S}$;%
\footnote{This is similar to what happens for $AdS_5\times S^5$, where the 16-dimensional representation of $\mathfrak{su}(2|2)^{\oplus2}$ for fundamental excitations can be written as tensor product of two 4-dimensional representations of $\mathfrak{su}(2|2)$.}
at the algebra level, this is tantamount to writing
\begin{equation}
    \gen{Q}^1 = \gen{Q}\otimes \gen{1}\,,\qquad
    \gen{Q}^2 = \gen{1}\otimes \gen{Q}\,,\qquad
    \dots\,,
\end{equation}
see \cite{Lloyd:2014bsa} for details on the decompositions into two-dimensional representations. The two-dimensional representations have the form  discussed in~\cite{Borsato:2012ud}, where the braiding factors for left-left, left-right, etc., representations was originally worked out.

Having understood that, in the limit $w\to\infty$, the $\mathcal{Z^\pm}$ markers yield braiding factors but do not result in new states, it is easy to understand their effect on the S~matrix. Because the S~matrix (up to a normalisation) is fixed by the commutation relations with the supercharges in the two-particle representations, it knows about the non-trivial coproduct. This choice of co-product can be modified, however, by means of a change of the two-particle basis (as we discussed above). In this way, the S~matrix can be translated to the string frame (or, should one want, to other coproducts). The change of the two-particle basis has to be properly taken into account when writing the Bethe-Yang equations of the model, of course, by properly redefining the volume.%
\footnote{In a sense, the whole point of the string frame versus spin-chain frame discussion is that the natural notions of volume are different in the spin-chain and on the string worldsheet; in one case it is the R-charge \textit{of the vacuum}, in the other it is the charge under $\gen{R}$ \textit{of the state}. This has been discussed in a context closely related to this one in~\cite{Dei:2018mfl}.}

Finally, the fact that the presence of the  $\mathcal{Z^\pm}$ markers is particularly evident in the chiral-antichiral (or left-right) sector, as highlighted already in eqs.~(3.10) and~(3.13) in~\cite{Borsato:2013qpa}, is simply due to the fact that there are processes which preserve $\gen{H},\gen{M},\gen{B}$ without preserving $\gen{R}$. This just cannot happen when scattering two isomorphic representations. If one would like to keep track of the charge under $\gen{R}$ and balance it between the left and right side of the equation, it is necessary to include $\mathcal{Z^\pm}$ markers, keeping in mind however that they are irrelevant at $w\to\infty$, due to the identifications of the type of eqs.~\eqref{eq:zmarkerid1}, \eqref{eq:zmarkerid2}, and \eqref{eq:zmarkerid3}. Indeed, if one accepts such identifications at the level of the algebra (which is necessary to construct a consistent representation), there is no reason not to do so at the level of the S~matrix.

\subsection{On the ``matching'' with the pp-wave spectrum}
It is also worth discussing the matching with the pp-wave spectrum on which the authors of~\cite{Gaberdiel:2023lco} put some emphasis.%
\footnote{We also thank the anonymous referee for underscoring the opportunity of further discussing this point.}
The analysis of Berenstein, Maldacena and Nastase (BMN)~\cite{Berenstein:2002jq} suggested that it was possible to match excitations in a limit of $AdS_5\times S^5$ strings (at large tension~$T\sim \sqrt{\lambda}$ and expanded around a null geodesic of large angular momentum~$J$, which yields the lightcone gauge-fixed string on the pp-wave geometry) with ``magnons'' of $\mathcal{N}=4$ SYM in a double-scaling limit where one is left with a coupling constant~$\lambda'=\lambda/J^2$ (where, in $\mathcal{N}=4$ SYM, $\lambda$ is the 't Hooft coupling and $J$ the R-charge, and both are large). Ref.~\cite{Berenstein:2002jq} also suggested that similar constructions may apply to other AdS/CFT setups, including $AdS_3\times S^3\times T^4$, though initially a great deal of attention was devoted to the case of $\mathcal{N}=4$ SYM. It was relatively soon understood~\cite{Serban:2004jf,Beisert:2004hm} that the matching in~$\lambda'$ was not exact, see also~\cite{Sieg:2010jt}. In hindsight, the underlying integrable structure and the large amount of supersymmetry are the reasons for the matching, and it just so happens that the first few orders of the~$\lambda'$ expansion match, due to the particular form of the S~matrix~\cite{Beisert:2004hm} and of the simple form of the relation between the 't Hooft coupling and the string tension.%
\footnote{Things are much more complicated in $AdS_4$/$CFT_3$, where the relation between the string tension and the Chern-Simons coupling is quite nontrivial, see~\cite{Klose:2010ki}.}
It is by now understood that the near-pp-wave expansion may be used as a test of \textit{e.g.}\ integrability conjectures only in the large-tension (and large-$J$) regime.

In the case of $AdS_3\times S^3\times T^4$ the near-pp-wave expansion of the string model has been studied extensively, especially with reference to the S~matrix in the integrability literature. When dealing with the pp-wave geometry proper, the light-cone gauge-fixed theory is free. The spectrum is given by oscillators with dispersion%
\footnote{This was essentially derived in the original work~\cite{Berenstein:2002jq}; for a more detailed analysis see also~\cite{Hoare:2013pma} for the case $\mu=\pm1$ and see~\cite{Lloyd:2014bsa} for general~$\mu$.}
\begin{equation}
\label{eq:dispersionPP}
    H(\mathsf{p},\mu) = \sqrt{\mathsf{p}^2+2q\mu \mathsf{p}+\mu^2}\,,
\end{equation}
where $0\leq q\leq  1$ is a property of the background and $\mu=0,\pm1$. 
The worldsheet momentum $\mathsf{p}$ is quantised in units of $T/w$ where $w$ is the $R$-charge of the lightcone gauge and $T$ is the tension.%
\footnote{We call tension~$T$ the dimensionless quantity~$T=R_{AdS}^2/(2\pi\alpha')$ where $R$ is the $AdS_3$ radius and $\alpha'$ is the string slope function.}
The pp-wave worldsheet Hamiltonian is the leading-order term of an expansion of the light-cone gauge-fixed  worldsheet model of $AdS_3\times S^3\times T^4$ strings, in the limit in which the string tension is large, the $\mathbf{R}$-charge is large, and their ratio is fixed. In this way, one can find that the ``free'' pp-wave picture is corrected by interactions, and they can be used to derive the tree-level near-pp-wave S~matrix (as discussed in~\cite{Hoare:2013pma} for the mixed-flux background and $\mu=\pm1$ modes). In fact, one can do more and even consider loops, which correspond to sub-leading orders in the near-pp-wave expansion, see e.g.~\cite{Sundin:2016gqe} and references therein; all these computations had been originally considered in the case of $AdS_5\times S^5$, see~\cite{Klose:2006zd}.

For the mixed-flux $AdS_3\times S^3\times T^4$ with parameters $h>0$ and $k\in\mathbb{N}$, the near-pp-wave limit requires the tension~$T$,
\begin{equation}
    T=\sqrt{\frac{k^2}{4\pi^2}+h^2}\,,
\end{equation}
to be large; the lightcone momentum (the eigenvalue~$w$ of~$\gen{R}$) is also large, in such a way that $T/w$ is fixed, and the worldsheet momentum of pp-wave excitation is also small. This is not the \textit{decompactification} limit which we have discussed so far (both in the orbifold or on the worldsheet) where $T$ is finite, $p$ is finite, and only $w$ is large (see also~\cite{Arutyunov:2009ga} for a pedagogical explanation).
To recover the pp-wave dispersion~\eqref{eq:dispersionPP} from the one which we obtained by symmetry in decompactification limit~\eqref{eq:dispersion} we can set~\cite{Hoare:2013pma,Lloyd:2014bsa}
\begin{equation}
\label{eq:pplimit}
    k = 2\pi q\, T,\quad
    h = \sqrt{1-q^2}\,T\,,\quad
    p = \frac{\mathsf{p}}{T}\,,\qquad T=\sqrt{\frac{k^2}{4\pi^2}+h^2}\gg1,\quad \mathsf{p},\,q~\text{fixed}.
\end{equation}
where $q$ and $\mathsf{p}$ remain finite as $T\to\infty$.
Because the pp-wave limit requires the tension to be infinite, this limit gives only very limited insight into finite-$k$ backgrounds. Indeed, in the pp-wave  limit such backgrounds have $q=0$ and are indistinguishable from the pure-RR $k=0$ background as $T\to\infty$.

This general setup is at odds with the discussion of the BMN limit of~\cite{Gaberdiel:2023lco}. There, the authors first of all restrict to the case $\mu=-1$; this is because even if they are considering torus modes, which have $\mu=0$, in effect they have decided to expand their momentum as 
\begin{equation}
\label{eq:GGNstrangeppwave}
    p = 2\pi-\frac{\mathsf{p}}{T}\,,\qquad \mathsf{p}\ll1,\qquad T~\text{fixed, with}\quad k=1,\quad h\ll1.
\end{equation}
This is not the standard near-pp-wave expansion; first of all the worldsheet momentum should be expanded around zero. In this case, the net effect of this choice looks like shifting $\mu$, due to the periodicity~\eqref{eq:kperiodicity}. This  explains why, at the end of the day, the authors find a formal matching with the $\mu=-1$ dispersion.%
\footnote{One could also just study modes with $\mu=\pm1$ in the usual near-pp-wave expansion, and expand the momentum around zero; see Table~\ref{tab:munotzero} for the matching of the $\mu=\pm1$  modes with the modes of the currents.}
Secondly and more importantly, the authors work at fixed tension (fixed $k=1$, and finite~$h$) while taking $\mathsf{p}\ll1$ as in~\eqref{eq:GGNstrangeppwave}. Hence, they are just considering a particular expansion of decompactified light-cone gauge-fixed model; again it is different from the pp-wave expansion where the tension must become large. They find $q=(1+4\pi^2h^2)^{-1/2}$, which is what could be found just from inverting the well-known relations~\eqref{eq:pplimit} at fixed $h,k$.%
\footnote{The same result would have followed by considering~\eqref{eq:dispersion}, expanding the momentum as $p=2\pi M-\mathsf{p}/T$ and comparing with the pp-wave dispersion for a given~$\mu$. This is  just a consequence of~\eqref{eq:kperiodicity} and that the expression under the square root is quadratic at small~$\mathsf{p}$. Each coefficient of the second order polynomial may be fixed just by allowing for sufficiently many parameters in the matching (the constant piece gives $M=-\mu$, the linear piece fixes units in which the momentum~$\mathsf{p}$ is quantised, and the quadratic term fixes~$q$).}
 In this sense, this matching does not contain any information aside from that coming from~\eqref{eq:pplimit}, which is however not valid in the small-tension regime of the orbifold theory. The symmetric-orbifold results should instead be matched with the decompactified string for fixed $h,k$ or, at most, with the pp-wave background at $T\to\infty$ (which necessarily means~$q=0$) if one takes the limit of a very strong deformation.

\subsection{On bound states in the \texorpdfstring{$h>0$}{h>0} model}
\label{sec:boundstateshgrt0}
We finally would like to comment on the bound-state discussion of~\cite{Gaberdiel:2023lco} in more detail.%
\footnote{We thank the anonymous referee for raising this matter.}
The S-matrix of~\cite{Lloyd:2014bsa} can, when expressed in terms of the Zhukovsky variables $x^{\pm}_{1}$ and $x^{\pm}_{2}$ develop poles or zeros which reduce its rank; this happens for instance in the left-left sector. In fact, the condition for bound states is, in terms of the Zhukovsky variables, exactly the same as for the pure-Ramond-Ramond S~matrix~\cite{Borsato:2012ud} and it reads%
\footnote{These bound states have been studied for the RR case in~\cite{Borsato:2013hoa,Frolov:2021fmj,Seibold:2022mgg} and for the mixed-flux case in~\cite{Hoare:2013lja,Frolov:2023lwd,OhlssonSax:2023qrk}. Note also that this is formally the same bound-state condition of $AdS_5\times S^5$.}
\begin{equation}
\label{eq:boundstatecondition}
    x^{+}_1 = x^{-}_2\,.
\end{equation}
Under this condition the tensor product of two short representation of ``mass'' and momentum $(\mu_1, p_1)$ and $(\mu_2, p_2)$ becomes reducible, and it yields a short representation of mass and momentum $(\mu_1+\mu_2, p_1+p_2)$~\cite{Borsato:2013hoa,Frolov:2021fmj,Seibold:2022mgg}. While this is true algebraically, regardless of the physical interpretation of the Zhukovsky variables, it is however worth noting that the equation~\eqref{eq:boundstatecondition} \textit{may not have a physical solution}, and that the bound state representation \textit{may not be distinct from a fundamental-particle representation}.

For instance, in the pure-RR case, where $k=0$ and $h>0$, this equation has solution when $x_j^\pm= x^{\pm}(p_j, \mu_j)$ with $\mu_1=\mu_2=1$ (or $\mu_1=\mu_2=-1$). The solution, in momentum space, can be thought as taking the form%
\footnote{%
In practice, for the RR model it is actually more appropriate to discuss bound states in terms of the $u$-rapidity, $u=x^\pm+1/x^{\pm}\mp 2i/h$, see \textit{e.g.}~\cite{Frolov:2021fmj}. This is very similar to what happens for $AdS_5\times S^5$ strings. Because the $u$-plane has cuts, the function $f(p,h)$ needs not be real in general.
}
\begin{equation}
\label{eq:boundstatesol}
    p_1 = \frac{p}{2}+i f(p,h)\,,\qquad
    p_2 = \frac{p}{2}-i f(p,h)\,,
\end{equation}
for some appropriate function $f(p,h)$, so that the total energy is real and positive (and the total momentum is real and equal to~$p$). However, this equation has no solution if $\mu_1=0$, or $\mu_2=0$, or both. In the RR case, it so happens that all representations with $\mu=2,3,4\dots$ are distinct from each other, but this is not necessarily the case (in fact, in relativistic integrable models typically the bound-state fusion procedure closes onto itself after finitely-many steps). 

To begin with one must carefully study which configurations of particles allow for a bound-state solution. In principle this requires a detailed discussion of the \textit{physical region} of the model, along the lines of~\cite{Arutyunov:2007tc}, establishing which values of \textit{complex momentum} are allowed in the model. For instance, in the case of a relativistic model, bound states are physical when the difference of the  relativistic rapidities of the particles%
\footnote{Recall that for a relativistic model with particles of mass $m$, $H=m\cosh\vartheta$ and $p=m\sinh\vartheta$.}
$\vartheta_{21}=\vartheta_2-\vartheta_1$ is lying in the physical strip $0<\text{Im}[\vartheta_{21}]<\pi$.
In the case at hand, for  $h>0$ the model is non-relativistic. It appears that it easiest to describe the physical region of the model in terms of deformed $u$-variables (first introduced in~\cite{Hoare:2013lja}),
\begin{equation}
\label{eq:uplane}
    u_{\scriptscriptstyle\text{L}}(x) = x+\frac{1}{x}-\frac{k}{\pi h}\ln(x)\,,\qquad
    u_{\scriptscriptstyle\text{R}}(x) = x+\frac{1}{x}+\frac{k}{\pi h}\ln(x)\,,
\end{equation}
whereby, using the inverse functions $x_{\scriptscriptstyle\text{L},\text{R}}(u)$ we have $x^\pm_{{\scriptscriptstyle\text{L}},j} = x_{\scriptscriptstyle\text{L}}(u\pm \tfrac{i h}{2}|\mu_j|)$ and $x^\pm_{{\scriptscriptstyle\text{R}},j} = x_{\scriptscriptstyle\text{R}}(u_j\pm \tfrac{i h}{2}|\mu_j|)$. The map $x_{\scriptscriptstyle\text{L},\text{R}}(u)$ has three branch cuts, two of which are related to the branch cut of $\ln(x)$, while the remaining is of the square root type, see~\cite{Frolov:2023lwd}.
The definition of a physical region for complex momenta, or complex~$u$, can be done in terms of the $u$-plane with cuts~\cite{Frolov:2023lwd}. This definition goes hand in hand with the discussion of crossing symmetry (which requires an analytic continuation to the anti-particle region through complex momenta) and the construction of a crossing-invariant dressing factor, which was done very recently~\cite{Frolov:2024pkz}.%
\footnote{The analytic structure of this model was also recently discussed in~\cite{OhlssonSax:2023qrk}, but their discussion of the physical region is not entirely clear to us.}
Without delving too deep in the analytic properties of the model, we would like to briefly remark on the discussion of bound states presented in~\cite{Gaberdiel:2023lco}.

In \cite{Gaberdiel:2023lco} the authors notice the presence of a bound state pole for the ``LL'' sector of the S-matrix. In the notation of the existing integrability literature, they consider the S~matrix of~\cite{Lloyd:2014bsa} with $k=1$, $h>0$ they take two particles with $\mu_1=\mu_2=0$ and $0<\text{Re}[p_j]<2\pi$ (somewhat confusingly, they assign to massless particles bound state number $Q=1$, whereas it should really be $Q=0$). Because the S~matrix for the RR model and for the mixed-flux model have the same form in terms of Zhukovski variables, the bound-state pole noticed by~\cite{Gaberdiel:2023lco} is  formally exactly the same as in~\cite{Borsato:2012ud,Borsato:2013hoa,Frolov:2021fmj,Seibold:2022mgg}, when expressed in terms of Zhukovsky variables. The condition reads like in eq.~\eqref{eq:boundstatecondition}. Because the analysis of~\cite{Gaberdiel:2023lco}  does not considers whether a physical solution of~\eqref{eq:boundstatecondition} exists, or what form the bound state representation takes, they seem to find the same structure of bound states as in the Ramond-Ramond model or in $AdS_5\times S^5$ --- infinitely many distinct bound states with bound-state number $Q=1,2,3,\dots$ .

Without having delving into the discussion of the physical regions in terms of the $u$-rapidity~\cite{Frolov:2023lwd}, it is easy to see that \eqref{eq:boundstatecondition} does not always admit a solution for the mixed-flux model. To this end, we can study it in terms of the momenta and see (numerically) that for small $p>0$ no  solution like~\eqref{eq:boundstatesol} exists. It is necessary to take $p$ to be sufficiently large for such solutions to exist, and in particular solutions of the type~\eqref{eq:boundstatesol} appear for $p>2\pi$ for~$h\ll1$.
Moreover, it is also easy to see that the bound-state representations that one may construct are not all inequivalent (unlike what is the case in the pure-RR model).

This fits with the discussion of~\cite{Frolov:2023lwd,Frolov:2024pkz}, which we will briefly repeat. In fact, let us consider general~$k$, and specialise it to $k=1$ later. There are two equivalent pictures. In the one picture, we allow the real part of the momentum~$p$ to take any value; in the context of the $u$-plane, this means gluing several sheets through the cut related to $\ln(x)$ in~\eqref{eq:uplane} (recall that $p=-i\ln x^++i\ln x^-$). At the level of the S~matrix, this requires an analytic continuation which is trivial for the matrix part of the S~matrix, but nontrivial for the dressing factor (because the dressing factor is not a meromorphic function of $x^\pm$, see~\cite{Frolov:2024pkz}). This allows us to define the S~matrix on a covering of the $u$-plane, similar to what considered in~\cite{OhlssonSax:2023qrk}. The particles created in this way can in general form bound states, but only $(k-1)$ distinct bound-state representations exist. The $k$-th bound state representation is isomorphic to the original fundamental particle representation, owing to~\eqref{eq:kperiodicity}, because the central charges $\gen{M}$ ,$\gen{H}$, $\gen{B}$ take the same value and they are the only thing in the decompactification limit $w\to\infty$ (where the S~matrix is defined).%
\footnote{%
Strictly speaking, one should also consider the Fermion grading of the resulting module, \textit{i.e.}~the fermion sign $(-1)^F$ of the highest weight state of the module; this is discussed in~\cite{Frolov:2023lwd}. Additionally, not all $(k-1)$ bound-state representations should be truly distinct: half of them should be related to the other half by charge conjugation.  
}
As remarked in Section~\ref{sec:orbifold:boundstates}, the charge $\gen{R}$ is not the same, but $\gen{R}$ decouples at $w\to\infty$ (and it is reinstated in the Bethe-Yang equations by correctly identifying the reference state). Therefore, the ``fused'' S~matrix of the $k$-th bound state should be related to the S~matrix analytically continued at $p\to p+2\pi$ through the cut of $\ln(x)$. This is a tautology for the matrix part of the S~matrix (and it can be easily checked); it is however a non-trivial constraint on the analytic structure of the dressing factor, on which we plan to return in future work. For the case of $k=1$, it means that for the purpose of constructing the S-matrix it is sufficient to continue the momentum to arbitrary real values. The alternative point of view is to insist that the real part of the momentum should be bounded by~$2\pi$, or in terms of $u$ that only the fundamental Riemann sheet is part of the physical region. In this case it natural to describe particles in regions with  larger and larger values of~$|\text{Re}[p]|$ as arising from bound states with larger and larger values of~$|\mu|$, again using~\eqref{eq:kperiodicity}. It is perhaps appealing that in this way one can make contact with the composite nature of the particles with $|\mu|=1$, which can be identified with modes of the current as identified in Table~\ref{tab:munotzero} with mode numbers suitably constrained to a region of size~$w$. However, this is not particularly illuminating when one considers even larger values of $|\text{Re}[p]|$ or $|\mu|$.

We conclude this brief discussion by emphasising once more that one can either consider any real value of~$p$ (and then only $(k-1)$ distinct bound-state representations exist) or restrict $p$ to a fundamental region of real range~$2\pi$, i.e.\ restrict ~$u$ to a fundamental sheet (and then bound states should be considered, having care of correctly identifying the physical region for complex momenta). However, one should not do both (consider any~$p$ and count any bound state representations) as this  would result in over-counting. In the particular case of $k=1$, this can also be immediately seen by continuity when one takes~$h\to0$.

\section*{Acknowledgements}
A.S.~acknowledges discussions on this topic with Matthias Gaberdiel during the 2022 workshop ``Integrability in String, Field, and Condensed Matter Theory'' at Kavli Institute for Theoretical Physics.
A.S.~acknowledges support from the European Union --
NextGenerationEU, from the program STARS@UNIPD, under project ``Exact-Holography -- A new exact approach to holography: harnessing the power of string theory, conformal field theory, and integrable models'', from the PRIN Project n.~2022ABPBEY, ``Understanding quantum field theory through its deformations'', and from the CARIPLO Foundation ``Supporto ai giovani talenti italiani nelle competizioni dell'European Research Council'' grant n.~2022-1886 ``Nuove basi per la teoria delle stringhe''.
A.S.~also thanks the MATRIX Institute in Creswick \& Melbourne, Australia, for support through a MATRIX Simons fellowship during the preparation of this work, and in conjunction with the MATRIX program ``New Deformations of Quantum Field and Gravity Theories''. 

\bibliographystyle{JHEP}
\bibliography{refs.bib}
\end{document}